\documentclass[fleqn,usenatbib]{mnras}

\usepackage{newtxtext,newtxmath}

\usepackage[T1]{fontenc}

\DeclareRobustCommand{\VAN}[3]{#2}
\let\VANthebibliography\thebibliography
\def\thebibliography{\DeclareRobustCommand{\VAN}[3]{##3}\VANthebibliography}

\usepackage{graphicx}	
\usepackage{amsmath}	
\usepackage{amssymb}	

\usepackage{xcolor}
\definecolor{orange}{rgb}{1.0, 0.5, 0.0}

\usepackage{flexisym}
\usepackage{breqn}

\newcommand{\erm}[1]{\textcolor{black}{#1}}

\definecolor{fixred}{rgb}{0.85,0.0,0.0}
\newcommand{\corr}[1]{\textcolor{black}{#1}}

\title[Modeling general-relativistic plasmas]{Modeling general-relativistic
plasmas with collisionless moments and dissipative two-fluid magnetohydrodynamics}

\author[Elias R. Most et al.]{ Elias R. Most,$^{1,2,3}$
\thanks{emost@princeton.edu} Jorge Noronha,$^{4}$ and Alexander A.  Philippov$^{5,6}$
\\
$^{1}${Princeton Center for Theoretical Science, Princeton University,
Princeton, NJ 08544, USA}\\
$^{2}${Princeton Gravity Initiative, Princeton University, Princeton, NJ
08544, USA}\\
$^{3}${School of Natural Sciences, Institute for Advanced Study, Princeton,
NJ 08540, USA}\\
$^{4}${Illinois Center for Advanced Studies of the Universe, Department of Physics,
University of Illinois at Urbana-Champaign, Urbana, IL 61801, USA}\\
$^{5}${Center for Computational Astrophysics, Flatiron Institute,
Simons Foundation, New York, NY 10010, USA}\\
$^{6}${Department of Physics, University of Maryland, College Park, MD 20742, USA}
}

\date{Accepted XXX. Received YYY; in original form ZZZ}

\pubyear{2021}

\begin{document}
\label{firstpage}
\pagerange{\pageref{firstpage}--\pageref{lastpage}}
\maketitle

\begin{abstract}
Relativistic plasmas are central to the study of black hole accretion,
jet physics, neutron star mergers, and compact object magnetospheres. 
Despite the need to accurately capture the dynamics of these plasmas and the implications for relativistic transients, 
their fluid modeling is typically done using a number of (overly) simplifying assumptions, which do not hold in general.
This is especially true when the mean free path in the plasma is large compared to the system size, and kinetic effects start to become important.
Going beyond common approaches used in the literature, we describe a fully relativistic covariant 14-moment based two-fluid system appropriate for the study of electron-ion or electron-positron plasmas. This generalized Israel-Stewart-like system of equations of motion is obtained directly from the relativistic 
Boltzmann-Vlasov equation. This new formulation can account for non-ideal 
effects, such as anisotropic pressures and heat fluxes, not present in previous formulations of two-fluid magnetohydrodynamics.
We show that a relativistic two-fluid plasma can be recast as a single
fluid coupled to electromagnetic fields with (potentially large)
out-of-equilibrium corrections. We keep all electron degrees of
freedom, which provide self-consistent evolution equations for electron
temperature and momentum. The out-of-equilibrium corrections take the form of a collisional
14-moment closure previously described in the context of viscous single
fluids. The framework also describes Braginskii-like
two-fluid magnetohydrodynamics applicable to weakly
collisional plasmas inside accretion disks.

\end{abstract}




\section{Introduction}
\label{sec:intro}
Relativistic flows are ubiquitous in high-energy astrophysics.
Accretion onto black holes can power electromagnetic transients
associated with the formation of a highly relativistic jet, e.g. 
short gamma-ray bursts in the case of stellar mass black holes formed in 
neutron star collisions \citep{LIGOScientific:2017zic}, 
or in supermassive black hole accretion \citep{Prieto:2015efa}.
Supermassive black holes, such as M87, feature intriguing
flaring activity \citep{Berge:2006uls}, which has recently been related to large scale
reconnection happening in the vicinity of the black hole \citep{2021MNRAS.502.2023P,2020ApJ...900..100R, 2020MNRAS.495.1549N,Ripperda:2021zpn}.
Such reconnection events are, however, not limited to black holes, but will
also occur { in over twisted neutron star magnetospheres} \citep{Parfrey:2013gza}.
For sufficiently large magnetic fields, the rotation of the star or orbital
motion, if it is in a close binary, can create electric fields close to the
surface that are large enough to accelerate surfaces charges to the
pair-creation limit \citep{Goldreich:1969sb}. Such an electron-positron plasma, also present around
supermassive black holes \citep{Moscibrodzka:2009gw,Wong:2020mif}, 
will have a large mean-free-path compared to the
size of the compact object, and will therefore be collisionless \citep{2020PhRvL.124n5101C,2021PhRvL.127e5101B}.
Reconnection events around black holes and neutron stars can trigger 
relativistic flaring events. These flares in turn can power associated
X-ray and radio transients, and have been proposed as a mechanism to explain Fast Radio Bursts \citep{Lyubarsky:2014jta,Beloborodov:2017juh,2019MNRAS.485.4091M,Beloborodov:2019wex,Yuan:2020ayo,Lyubarsky:2020qbp}, giant pulses in the Crab pulsar \citep{Lyubarsky:2018vrk,Philippov:2019qud}, 
and electromagnetic precursors in neutron star collisions \citep{Most:2020ami,2020arXiv201107310B}.

Collisionless plasmas are also highly relevant for the dynamics of
accretion disks around supermassive black holes. Temperature gradients can drive effective
heat fluxes, and fluctuations in the magnetic field {lead to pressure} anisotropies \citep{Chandra:2015iza,Foucart:2015cws}, which could
potentially impact the global dynamics of supermassive black-hole accretion
flows \citep{Foucart:2017axc}. Interestingly, regions where the
magnetic pressure is small compared to the thermal pressure, become mirror-
and firehose- unstable, increasing the effective collisionality of the
plasma \citep{Kunz:2014qha}. Such plasmas are then effectively weakly collisional \citep{1965RvPP....1..205B}.
Yet, they still require an accurate treatments of both electrons and ions in
order to correctly capture the heating of electrons, producing observable
signatures for the accretion flows, as observed by the Event Horizon
Telescope \citep{EventHorizonTelescope:2019dse}. 

While there is an urgent need for the accurate modeling of relativistic
plasmas in (weakly) collisionless scenarios, current numerical models are
severely limited in several aspects. Firstly, in collisionless plasmas that are directly governed by solutions of the Boltzmann-Vlasov equation the characteristic rate at
which magnetic fields reconnect is roughly given by $v_{\rm inflow}/v_A \approx
0.1$, where $v_A$ is the Alfven speed and $v_{\rm inflow}$ is the inflow speed
of reconnecting magnetic field lines into the current sheet (e.g., \citealt{2014ApJ...783L..21S}).
However, most simulations using general-relativistic magnetohydrodynamics \citep{EventHorizonTelescope:2019pcy}
formally appropriate only for collisional plasmas, reproduce a reconnection
rate that is too low by up to a factor 10 \citep{2021PhRvL.127e5101B}. This has important
consequences for setting the right time scale in the reconnection process
and connecting it to observations \citep{Ripperda:2021zpn}.

Secondly, in such simulations magnetic reconnection is typically captured only in
terms of a scalar resistivity (e.g. \citealt{Palenzuela:2008sf,Ripperda:2019lsi}),
or even just grid dissipation {in ideal MHD simulations}. Detailed investigations in the context
of space-physics plasmas indicate, however, that anisotropic pressure
contributions to Ohm's law might be the main driver of reconnection in
collisionless systems \citep{Bessho:2005zz,2015PhPl...22a2108W}.

Thirdly, effective collisionalities leading to heat fluxes are typically
not accounted for in simulations of black hole accretion, see \citealt{{Chandra:2015iza,Foucart:2015cws}} for notable exceptions. Similarly,
a consistent evolution of the electron temperature in a relativistic plasma
is largely unavailable, with approximations being currently in use
\citep{Ressler:2015ipa}. These require formulations that truly model the plasma as consisting of  
two interacting fluids.

Finally, we remark that a consistent relativistic description of viscous fluids is a complex task (see, e.g., review \citealt{Romatschke:2017ejr}), with standard formulations e.g. \citep{Eckart:1940te} displaying issues with causality and stability \cite{Hiscock:1985zz}. Designing schemes that are causal, stable, and strongly hyperbolic (well-posed)   
is a highly nontrivial feat even for purely viscous hydrodynamical descriptions, although those properties have been recently established in recent years \citep{Bemfica:2017wps,Bemfica:2019knx,Bemfica:2019cop,BDRS_2019, Bemfica:2020xym,Bemfica:2020zjp,Bemfica:2020gcl}. On the other hand,  magneto-hydrodynamic formulations can also face a number of shortcomings, including the lack of strong hyperbolicity \citep{Schoepe:2017cvt}.

In this work, we introduce a new description appropriate for relativistic
electron-ion and electron-positron plasmas, which goes beyond traditional
approaches currently in use. Before giving a detailed description of our
method, we review current approaches used in the literature, carefully
highlighting which physics they miss out, in the context of the discussion above.

In effective single fluid approaches, several studies have 
incorporated scalar resistivity in relativistic MHD (e.g.
\citealt{Palenzuela:2008sf,Zenitani:2010vq,Bucciantini:2012sm,
Dionysopoulou:2012zv,Qian:2016lyn,Ripperda:2019lsi,Wright:2019blb}).
The incorporation of scalar resistivity is done using a Newtonian-type
Ohm's law, which inevitably breaks strong hyperbolicity of the system
\citep{Schoepe:2017cvt}.
Going beyond simple scalar resistivity, several works have suggested the
inclusion of the Hall \citep{Zanotti:2011sv} or sub-grid dynamo effects
\citep{DelZanna:2018dyb,Tomei:2019zpj,Shibata:2021xmo}.
When modeling neutron matter, where weak interactions are relevant, it has
also been suggested to add composition-dependent terms to Ohm's law 
\citep{Andersson:2021kfk}.
Critically, these approaches are only valid in the regime where the plasma
mean free path is short, i.e. where collisions dominate. 
As such, they do not capture collisionless reconnection correctly, and cannot account
for anisotropic pressure effects, neither in reconnection or in viscous angular momentum transport in disks.

Instead of starting from phenomenological approaches, it is also possible
to derive relativistic single-fluid dynamics beyond the
(magneto-)hydrodynamical limit from the Boltzmann(-Vlasov) equation. Based
on a moment reduction of the Boltzmann equation it is possible to derive
general equations for out-of-equilibrium single fluid dynamics for
non-resistive \citep{Denicol:2018rbw,Panda:2020zhr} and resistive
\citep{Denicol:2019iyh,Panda:2021pvq} plasmas.  In particular, the so-called 14-moment decomposition
can be thought of as a relativistic extension of Grad's 13-moment approach
\citep{Grad13}, which accounts for heat fluxes and anisotropic stresses.
Different from the original formulation of \citet{Israel:1979wp}, the
approach of \citet{Denicol:2012cn} allows for a systematic power counting expansion
in terms of inverse Reynolds and Knudsen numbers, whose asymptotic regime agrees with the
Chapman-Enskog expansion of the Boltzmann equation
\citep{chapman1990mathematical,Denicol:2012cn}.
The consistent derivation from 
the Boltzmann-Vlasov equations results in generalized Israel-Stewart-like equations
that are likely to possess a causal regime for some range of values of transport coefficients and dissipative fluxes  \citep{Biswas:2020rps}, and those equations can account for anisotropic pressures as 
well as heat conduction and bulk viscosities. However, they are formally only valid
in the regime of strong and weak collisionality.

While the studies above mainly concerned themselves with modeling a single
fluid, charged plasmas strictly require two individual components to
maintain charge quasi-neutrality. As such, modeling electron-ion plasmas in
black hole accretion problems or electron-positron plasmas, appropriate for
the study of neutron-star or black-hole magnetospheres, strictly requires
the inclusion of two (interacting) plasma components.  Several attempts to
model relativistic multi-fluids have been taken in the literature.  In the
case of modeling reconnection in electron-positron plasmas, early work had
tried to naively couple two separate perfect single fluids via a simple
momentum exchange term, in non-covariant
\citep{Zenitani:2009bj} and covariant form
\citep{Zenitani:2009di,Barkov:2013xga}. These approaches did not include further non-ideal
effects such as heat fluxes or anisotropic pressure contributions, which
might be significant for reconnection \citep{Bessho:2005zz,2015PhPl...22a2108W}.  On the other hand,
\citet{Koide:2009yx} proposed a simple set of equations to model
relativistic two-component plasmas, under the assumption that each plasma
can be modelled as a perfect fluid. While this model can account for the
interaction of electrons and ions through electromagnetic fields, it fails
to include heat fluxes and anisotropic electron pressure contributions to Ohm's
law. These equations miss out on important contributions to the
electron temperature evolution \citep{Ressler:2015ipa}, governing the
emission near accreting black holes \citep{Moscibrodzka:2009gw}.  Modeling
the interior of neutron stars, several variational formalisms (e.g.
\citealt{1991RSPSA.433...45C,Andersson:2013jga,Rau:2020wnv}) have been
proposed to model relativistic resistive and reactive plasmas
\citep{Andersson:2016wnc,Andersson:2016fva}.

\section{Motivation and Outline}
\erm{
Before we dive into the technical details of deriving a system of relativistic dissipative two-fluid magnetohydrodynamics equations, we want to sketch the overall picture and motivation of this work. To this end, we can summarize our discussion in Sec. \ref{sec:intro} as follows.
The goal of this paper is to describe a framework that can model dissipative effects in relativistic plasmas.
These may be important in dense plasmas found in neutron star mergers \citep{Alford:2017rxf,Most:2021zvc} or in accretion disks around supermassive black holes \citep{Foucart:2015cws,Foucart:2017axc}.
Moreover, the key effect of dissipation on a magnetized plasma is to allow for dissipation of electromagnetic energy, which happens in reconnecting current sheets (see, e.g., \citealt{2022arXiv220209004J} for a recent review).
This form of dissipation is crucial in understanding reconnection powered transient produced in compact object magnetospheres \citep{Beloborodov:2017juh,Lyubarsky:2020qbp}, precursor emission to compact object mergers \citep{Most:2020ami} and flaring activities around supermassive black holes \citep{Ripperda:2021zpn}.
Due to the collisionless nature of the electron-positron pair plasmas in these systems, their dynamics can in-principle be obtained ab-initio as solutions of the Vlasov-Boltzmann equation. However, the vast physical separation of scales of the plasma (e.g, electron skin depth and Larmor radius) and of the macroscopic system (e.g.,horizon size and orbital size) makes first principle approaches with correct separation of scales computationally unfeasible.} \erm{
Instead, what is needed from a practical point, is a framework that allows to simulate approximate scale separations, i.e., where the Larmor radius and skin depth enter as free parameters. Some of these scales (e.g., thermal Larmor radius) can also efficiently be overstepped numerically using implicit time integration methods \citep{Most:2021oha}, or only be locally resolved using adaptive mesh refinement techniques \citep{berger1989local}, which have been extensively used for fluid equations (see, e.g., \citealt{plewa2005adaptive}).
However, these advantages come at the cost of loosing the ability to correctly capture the saturation phase of plasma instabilities arising from kinetic physics (e.g., \citealt{PhysRevLett.2.83}).}\\

\erm{Leveraging these benefits of a fluid-like approach, we will formulate a two-fluid approach to dissipative magnetohydrodynamics. In order to systematically incorporate these corrections, we will utilize the 14-moment approach of \citet{Denicol:2012cn} (see also \citealt{Denicol:2018rbw, Denicol:2019iyh}), which is a second-order Müller-Israel-Stewart-type system \citep{Israel:1979wp}. Crucially, this system provides additional evolution equations for the anisotropic pressure $\pi^{\mu\nu}$ and heat fluxes $q^\mu$,}
\erm{
\begin{align}
    u^\alpha \nabla_\alpha q^{\mu} &= \ldots\,,\\
    u^\alpha \nabla_\alpha \pi^{\mu\nu} &= \ldots\,,
\end{align}
}\erm{
thereby making the system (likely) causal \citep{Bemfica:2020xym}. Here $u^\mu$ is the fluid four-velocity.
In addition, the inclusion of effects of dissipation on the electromagnetic sector will require an effective Ohm's law.
Schematically, this will take the following form \citep{Denicol:2019iyh}\,,
}\erm{
\begin{align}
    e^\mu = \eta_1 j^\mu + \eta_2 b^{\mu\nu} j_\nu + \alpha \nabla_\nu \pi^{\mu\nu} + \beta u^\nu \nabla_\nu j^\mu +\ldots.
    \label{eqn:Ohm}
\end{align}
}\erm{
Here, $e^\mu$ is the electric field as seen by an observer comoving with the fluid four-velocity $u^\mu$, $j^\mu$
is the dissipative part of the electric current, and $b^{\mu\nu}$ is the
antisymmetric tensor constructed from the comoving magnetic field, see
\eqref{eqn:b_projector}. Terms with $\eta_i$ coefficients are first-order in gradients.\footnote{\erm{We point out that in a two-component plasma gradients can either be on ion or electron scales, complicating the notion of \textit{the} gradients in the above expressions. For clarity, we will drop this distinction for the most part of this paper.}}
It is important to point out that the advection operator, $ u^\nu\nabla_\nu j^\mu$, on the right-hand side provides an effective time-evolution equation for the electric current. In this sense, the Ohm's law becomes dynamical and takes a similar form to the Müller-Israel-Stewart-type equations discussion above.
While the validity of such an approach might not be directly apparent, it has been extensively investigated both in the context of collisionless Newtonian plasmas \citep{2015PhPl...22a2108W}, as well as relativistic pair plasmas \citep{2012ApJ...750..129B,Liu:2014ada}.
By comparing with fully kinetic simulations, e.g., \citet{2012ApJ...750..129B} and \citet{Liu:2014ada} have directly shown that second order terms are crucial in relativistic reconnection processes.
Building by similar findings in electron-ion reconnection, Newtonian ten-moment closures have been developed \citep{2015PhPl...22a2108W,2020PhPl...27h2106N}.
Motivated by their success in describing dissipation in planetary magnetospheres \citep{2018JGRA..123.2815W,2019GeoRL..4611584D}, we will construct a 14-moment two-fluid scheme
featuring an effective Ohm's law of the form \eqref{eqn:Ohm}.
}
More specifically, we propose an alternative formulation for two-fluid
(dissipative) magnetohydrodynamics appropriate for the study of
collisionless, weakly collisional, and collisional resistive plasmas.
Building on recent progress in the moment expansion of the relativistic
Boltzmann-Vlasov equation \citep{Tinti:2018qfb,Denicol:2019iyh}, we first
derive a set of collisionless 14-moment equations. These can be seen as the
relativistic generalization of 10-moment approaches used in space-physics
\citep{2015PhPl...22a2108W}. The resulting equations are presented in Sec.\ \ref{sec:gkyell}.
In the second part of this work, we show how the 14-moment representation
of two fluids can be recast to resemble an effective single-fluid
description with dissipative out-of-equilibrium corrections.  These
equations, as presented in Sec.\ \ref{sec:twofluids}, themselves resemble
the 14-moment closures presented in Sec.\ \ref{sec:gkyell}.  Specializing on
the case of an electron-ion plasma, in Sec.\ \ref{sec:eI} we discuss
simplified cases appropriate for weakly collisional plasmas. Finally, we
make a connection to a resistive dissipative magnetohydrodynamics
description of a single-fluid plasma in Sec.\ \ref{sec:dissMHD}.
Throughout this work we adopt geometric units, $G=c=k_B=1$, and a mostly plus signature for the spacetime metric $g_{\mu\nu}$. Lorentz scalars constructed via the scalar product among vectors are denoted with a ``$\cdot$", i.e., $p_\mu u^\mu = p\cdot u$.

\section{Relativistic fluids}

In the following, we will give a brief description of a single relativistic fluid
described in terms of 14 moments. While this section will introduce all
basic concepts and expressions, they will lack the statement of a closure,
the simplest of which being an equation of state.
These will be discussed separately in Sec.\ \ref{sec:gkyell}. 

A relativistic fluid can be described in terms of a symmetric energy-momentum tensor $T_f^{\mu\nu}$ and a conserved number density
$N^\mu_f$, i.e.
\begin{align}
  \nabla_\mu N^\mu_f =0.
  \label{eqn:Ncons}
\end{align}
The fluid energy-momentum tensor $T_f^{\mu\nu}$ is also subject to total energy-momentum conservation, i.e. $\nabla_\mu T_f^{\mu\nu}=0$.
Introducing a characteristic timelike 4-velocity $u^\mu$ of the fluid (with $u_\mu u^\mu=-1$), we can split
$N_f^\mu$ in components along and across $u^\mu$, i.e.
\begin{align}
  N_f^\mu = n_f u^\mu + V_f^\mu,
  \label{eqn:Nf}
\end{align}
where $V_f^\mu$ is typically referred to as particle diffusion current, as
it refers to particle motion across the fluid current $n_f u^\mu$.
By construction $V_f^\mu$ is orthogonal to $u^\mu$, i.e.
\erm{
\begin{align}
  \Delta^{\mu}_\nu N_f^\nu :=& \left( \delta^{\mu}_{\nu} + u^\mu u_\nu \right)
  N_f^\nu=V_f^\mu,
  \label{eqn:Delta}\\
  V_f^\mu u_\mu =& 0
\end{align}
}
where we have introduced the fluid frame projector $\Delta_{\mu\nu}$.
While in many astrophysical situations the diffusion current $V_f=0$, 
we will see in Sec. \ref{sec:twofluids}, that this depends on the choice of the fluid hydrodynamic frame. 
If the fluid particles have characteristic charge $\mathfrak{q}$, we can
introduce an electric fluid current, proportional to the conserved number density
current $N^\mu_f$,
\begin{align}
  \mathcal{J}_f^\mu = \mathfrak{q} N_f^\mu.
  \label{eqn:Jf}
\end{align}

From a fundamental point of view, gravity couples energy and momentum 
to the curvature of spacetime.  Within the framework of general relativity,
the Einstein equations impose that 
\begin{align}
  G^{\mu\nu} = 8 \pi T^{\mu\nu}\,,
  \label{eqn:einstein}
\end{align}
where $G^{\mu\nu}$ is the Einstein tensor and  $T^{\mu\nu}$ is the energy-momentum
tensor encompassing all matter and non-gravitational fields, e.g.
electric and magnetic fields. 
As a consequence of the Bianchi identity,
$\nabla_\mu G^{\mu\nu} = 0$, the
total energy-momentum is conserved
\begin{align}
  \nabla_\mu T^{\mu\nu}=0.
  \label{eqn:T_total}
\end{align}
This implies that the ten independent components of the $T_f^{\mu\nu}$ and the four
components of the number density current, $N_f^{\mu}$, are the natural variables
describing a relativistic fluid. Those 14 degrees of freedom will be referred as the {\it 14-moments} in this work. Their construction in terms of kinetic theory will be discussed in Sec. \ref{sec:gkyell}.

Introducing an energy density $e$ and equilibrium pressure $P=P\left(
e,n,\dots \right)$, we can
decompose the energy-momentum tensor $T_f^{\mu\nu}$ as follows
\begin{align}
  T_f^{\mu\nu} = e u^\mu u^\nu + \left( P+\Pi \right) \Delta^{\mu\nu} +
  q^\mu u^\nu + q^\nu u^\mu + \pi^{\mu\nu}.
  \label{eqn:Tfsplit}
\end{align}
In doing so, we have introduced the viscous bulk scalar $\Pi$, the  anisotropic stress
tensor $\pi^{\mu\nu}$, and the energy diffusion 4-vector $q^\mu$. By construction these are
orthogonal to the fluid four-velocity, i.e.
\begin{align}
&  u_\mu q^\mu =0,\\
&  u_\mu \pi^{\mu\nu} =0, \label{eqn:pi}
\end{align}
and additionally $\pi^\mu_\mu =0$. 
It is important to stress that this decomposition is purely
algebraic. Any symmetric rank two tensor can be split along the direction
of an arbitrary timelike vector $u^\mu$ in this way. It will turn out to be beneficial to also introduce the enthalpy $h$ per particle of the fluid,
\begin{align}
    h = m \mathfrak{h} = \frac{e + P}{n}\,,
\end{align}
where $m$ is the mass of the fluid particle and $\mathfrak{h}$ is the specific enthalpy.

In equilibrium, the viscous, or out-of-equilibrium, parts vanish and the baryon current and energy
momentum current reduce to their equilibrium values, i.e.
\begin{align}
&  N^\mu_{f\,,\rm eq} := n_f u^\mu, \\
&  T^{\mu\nu}_{f\,,\rm eq} := e u^\mu u^\nu + P \Delta^{\mu\nu},
\label{eqn:Teq}
\end{align}
which implies
\begin{align}
  u_\nu T^{\mu\nu}_{f\,,\rm eq} = - e u^\mu = - \frac{e}{n_f}\erm{N^\mu_{f\,, \rm eq}}.
  \label{eqn:uTf}
\end{align}
This means that in equilibrium the flow of energy aligns with the flow of
particles.
On the other hand, in the presence of the viscous corrections, these
relations no longer hold.
This can be seen as follows. Projecting the energy momentum tensor onto the
fluid four-velocity, we find
\begin{align}
  -u_\mu T^{\mu\nu}_f = e u^{\corr{\nu}} + q^{\corr{\nu}}.
  \label{eqn:uTf2}
\end{align}
For non-vanishing energy diffusion $q^\mu$, the flow of energy and particles is no
longer aligned, since $q_\mu u^\mu=0$. The same is also true for the current,
see \eqref{eqn:Nf}.
The actual heat flux can be computed by removing the energy carried by particle diffusion,
\begin{align}
  \mathcal{Q}^\mu = q^\mu - h_0 V_f^\mu,
  \label{eqn:Q_tot}
\end{align}
where $h_0$ is the enthalpy per fluid particle in equilibrium. 

It is important to remark that we have 17 independent variables in the set  $\{e,u^\mu,\Pi,q^\mu,\pi^{\mu\nu},n_f,V_f^\mu\}$ introduced in \eqref{eqn:Nf} and \eqref{eqn:Tfsplit}. Thus, to recover a 14-variable description, choices have to be made for the fields that appear in these equations. Each choice for the variables used in the description of the energy-momentum tensor and the conserved current defines the so-called hydrodynamic frame \citep{Stewart:1972hg,Israel:1979wp,Tsumura:2006hnr,Kovtun:2012rj}, of which well-known choices are the Landau hydrodynamic frame \citep{LandauLifshitzFluids},
where $q^\mu=0$ and the flow of energy aligns with the fluid
four-velocity (no energy diffusion), and the Eckart frame \cite{Eckart:1940te}, where particle
diffusion is absent, i.e. $V_f^\mu=0$. There is an infinite number of possible hydrodynamic frames, and we refer the reader to Refs.\ \cite{Bemfica:2017wps,Kovtun:2019hdm,Bemfica:2019knx,Hoult:2020eho,Bemfica:2020zjp} for a discussion of such choices and their properties and consequences. For instance, one can rigorously prove \citep{ Bemfica:2017wps,Bemfica:2019knx,Bemfica:2019cop,BDRS_2019, Bemfica:2020xym,Bemfica:2020zjp,Bemfica:2020gcl} that first-order viscous hydrodynamic theories can be causal and stable when constructed using general hydrodynamic frames. In this work it will be most natural to consider the Eckart frame. However, for the sake of generality, for now we will not further specify our choice and keep both dissipative components, $V_f^\mu$ and $q^\mu$, in the equations.

Before moving on to describe a set of closure relations for the
out-of-equilibrium terms $\Pi, q^\mu, V_f^\mu$, and $\pi^{\mu\nu}$, we need to
introduce a description of electric and magnetic fields in general
spacetimes. This is done in the next section.

\subsection{Electromagnetic fields in relativity}

In covariant form, the electromagnetic field is described by a \erm{totally antisymmetric} field
strength tensor $F^{\mu\nu}$, and its dual ${^\ast}F^{\mu\nu} = \corr{\frac{1}{2}} \varepsilon^{\mu\nu\alpha\beta} F_{\alpha\beta}$.
The dynamics of the electric and magnetic fields is then governed by the
Maxwell equations
\begin{align}
&  \nabla_\mu ^{\ast}F^{\mu\nu}=0,\\
&  \nabla_\mu F^{\nu\mu} = 4\pi \mathcal{J}^\nu
  \label{eqn:maxwell}
\end{align}
where $\mathcal{J}^\nu$ is the total electric current.
Due to the antisymmetric property of the field strength tensor
$F^{\mu\nu}$, the Maxwell equations \eqref{eqn:maxwell} imply global conservation
of electric charge,
\begin{align}
  \nabla_\nu \mathcal{J}^\nu =0.
  \label{eqn:Jcons}
\end{align}

The electric and magnetic fields further give rise to an energy-momentum
tensor
\begin{align}
  T_{\rm EM}^{\mu\nu} = \frac{1}{4\pi}\left( F^{\mu\alpha} {F^\nu}_{\corr{\alpha}} -
  \frac{1}{4}g^{\mu\nu} F^{\alpha\beta} F_{\alpha\beta} \right),
  \label{eqn:TEM}
\end{align}
which obeys the following conservation law
\begin{align}
  \nabla_\mu T^{\mu\nu}_{\rm EM} = - F^{\nu\lambda} \mathcal{J}_\lambda.
  \label{eqn:TEMcons}
\end{align}

Using the fluid four-velocity $u^\mu$, we can project the field strength
tensor $F^{\mu\nu}$ and the corresponding electric and magnetic fields obtained in this
procedure are then those seen by an observer comoving with the fluid.
Thus, we can define the comoving
fields via 
\begin{align}
  F^{\mu\nu} u_\nu  &= e^\mu\,, \\
  {^\ast\!F}^{\mu\nu} u_\nu  &= b^\mu\,.
  \label{eqn:comoving_fields}
\end{align}
We can furthermore introduce the following tensors
\begin{align}\label{eqn:b_projector}
    b^{\mu\nu} &=  \varepsilon^{\mu\nu\alpha\beta} b_\alpha u_\beta\,, \\
    e^{\mu\nu} &=  \varepsilon^{\mu\nu\alpha\beta} e_\alpha u_\beta\,, 
\end{align}
which allow us to re-express the field strength tensor entirely in terms of the comoving
fields. It is then given by,
\begin{align}
  F^{\mu\nu} &= u^\mu e^\nu - u^\nu e^\mu + b^{\mu\nu},
  \label{eqn:Fmunu}\\
  ^\ast\!F^{\mu\nu} &= u^\mu b^\nu - u^\nu b^\mu - e^{\mu\nu}.
  \label{eqn:Fast}
\end{align}
The electromagnetic energy-momentum tensor \eqref{eqn:TEM} then 
takes the convenient form
\begin{dmath}
  4\pi T_{\rm EM}^{\mu\nu} = -\frac{1}{2} \left( b^2 + e^2 \right)
  g^{\mu\nu} + \left( b^2 + e^2 \right) \Delta^{\mu\nu}  - b^\mu b^\nu -
  e^\mu e^\nu \corr{+} e_\alpha u^\mu b^{\nu\alpha} \corr{+} e_\alpha u^\nu b^{\mu\alpha}.
  \label{eqn:TEM_fluid}
\end{dmath}

This completes our short review of the covariant formulation of Maxwell equations used in this work.

\section{Moment equations and closure relations}
\label{sec:gkyell}

Having discussed the general framework of relativistic fluids, we need to
specify closure relations for the out-of-equilibrium corrections.
From the fundamental point of view of kinetic theory, these should be
described as reductions of the Boltzmann-Vlasov equation.
Such an approach has been taken by \cite{Tinti:2018qfb}, which we summarize in the
following.

A microscopic kinetic theory is described in terms of a distribution function
$f\left( x^\mu, p^\nu \right)$, which describes a distribution
of particles at point $x^\mu$ with momenta $p^\nu$. Given a single species
of particles with mass $m$, we have the on-shell relation $p_\mu p^\mu =-m^2$.
This distribution function obeys the Boltzmann-Vlasov equation,
\begin{align}
  p^\mu \partial_\mu f + \left[\mathfrak{q} F^{\alpha\mu} p_\mu \corr{-}
  \Gamma^\alpha_{\mu\nu} p^\mu p^\nu \right] \partial_{p^\alpha} f =
  C\left[ f \right],
  \label{eqn:BoltzmannR}
\end{align}
where we have included couplings to electromagnetic fields and a suitable
collision operator $C\left[ f \right]$, see also \citet{Denicol:2019iyh}.
It should be pointed out that the appearance of the Christoffel symbol $\Gamma^\alpha_{\mu\nu}$ is associated with ensuring general covariance of the Boltzmann equation. In all what follows we will, without loss of generality, perform our calculations in flat Minkowski spacetime. The resulting equations are, however, manifestly covariant and correct also in general-relativistic spacetimes as long as long as the full covariant derviative is used.

In order to obtain a fluid description we need to remove the momentum
dependence, which is replaced by an average velocity $u^\mu$ describing the bulk motion of the fluid.
It turns out to be beneficial to systematically introduce an irreducible basis of
particle momenta relative to the fluid four-velocity $u^\mu$.
That is, we are going to assemble products of the form
$\left( -p\cdot u \right)$, $p^{<\mu>}$, $p^{<\mu>}p^{<\nu>}$,$\ldots$ ,
where we have introduced the notation $p^{<\mu>} = \Delta^\mu_\alpha
p^\alpha$, with $\Delta^\mu_\alpha= \delta^\mu_\alpha + u^\mu u_\alpha$. This basis can then be used to expand the distribution function $f$.
We further introduce the notion of a generalized covariant moment
\begin{align}
  f^{\mu_1\dots\mu_n}_r = \int_p \left( -p\cdot u \right)^r
  p^{\left<\mu_1\right>}\dots p^{\left<\mu_n\right>} f,
  \label{eqn:f}
\end{align}
for which the particle momentum dependencies have been integrated
out. Note that these moments are orthogonal to the 4-velocity. Since the momenta are on-shell, we have split out the momentum component along $u^\mu$, such that in the fluid rest-frame the Lorentz invariant measure in the integral can be written as 
\begin{align}
    \int_p := \int \frac{{\rm d}^3 p}{\corr{p^0}}\,,
\end{align}
where the integral runs over all spatial momenta.
Making use of the on-shell relation $p_\mu p^\mu = -m^2$, we can write
\begin{align}
  \left(p_\mu u^\mu\right)^2 = m^2 + p^{<\mu>} p_{<\mu>}\,,
  \label{eqn:pu}
\end{align}
and, thus, simplify the moments
\begin{align}
  f^{\mu_1\dots\mu_n}_r = \int_p \left(\sqrt{m^2 + p^{<\mu>} p_{<\mu>}}\right)^{r}
  p^{\left<\mu_1\right>}\dots p^{\left<\mu_n\right>} f,
  \label{eqn:f2}
\end{align}
having again used that the integral over the momentum is subject to
the on-shell constraint.
This decomposition also ensures that we can directly identify
the hydrodynamic degrees of freedom, as will be shown in the following.

In particular, the particle current $N^\mu_f$ and the energy momentum
tensor $T^{\mu\nu}_f$ can be expressed as
the first and second moment of the distribution function, respectively,
\begin{align}
  &N^{\mu}_f = \corr{\int_p} p^\mu f\,,\\
  &T^{\mu\nu}_f = \int_p p^\mu p^\nu f.
  \label{eqn:Tfromf}
\end{align}

Using the above definition, we can readily
identify some of the irreducible moments with our hydrodynamic variables,
\begin{align}
  n_f &= \corr{-}u_\mu N^\mu = \corr{f_1}\,,\\
  e &= u_\mu u_\nu T^{\mu\nu} = f_2\,,\\
  P+\Pi &= \frac{1}{3}\Delta_{\mu\nu} T^{\mu\nu} =\frac{1}{3} \Delta_{\mu\nu}f_0^{\mu\nu}\,,\\
  V_f^\mu &= \Delta_\nu^\mu N^\nu = \corr{f_0^\mu}\,,\\
  q^\mu &= -\Delta_\beta^\mu u_\alpha T^{\alpha\beta}= f_1^\mu \label{eqn:qmoment}\,,\\
  \pi^{\mu\nu} &= \Delta^{\mu\nu}_{\alpha\beta} T^{\alpha\beta} =
\erm{f_0^{<\mu\nu>}\,.}
\end{align}
In writing these expressions we have introduced the following notations for the projection of a vector and rank-two tensor into the fluid frame, $A^{<\mu>} = \Delta^\mu_\nu A^\nu$, $B^{<\mu\nu>}= \Delta^{\mu\nu}_{\alpha\beta} B^{\alpha \beta}$, where $\Delta^{\alpha\beta}_{\mu\nu} =
\left( \Delta^{\alpha}_{\mu}\Delta^{\beta}_{\nu} +
\Delta^{\beta}_{\mu}\Delta^{\alpha}_{\nu} \right)/2 -
\Delta^{\alpha\beta}\Delta_{\mu\nu}/3$ is the symmetric trace-free projector.
\erm{In writing the above, we have implicitly split the total pressure tensor},
{
\begin{align}
    \Pi^{\mu\nu}= \pi^{\mu\nu} +  \left(P+\Pi\right) \Delta^{\mu\nu} = f_0^{\mu\nu}\,,
\end{align}
}
\erm{into isotropic $\left(P+\Pi\right)$ and anisotropic $\left(\pi^{\mu\nu}\right)$ components.
It is important to understand the split of the isotropic pressure into an equilibrium, $P$, and dissipative, $\Pi$, part is
not always well defined. In the absence of an equilibrium equation of state, these two contributions cannot be separated.
To keep a consistent notation through this work, we retain this split explicitly in this section.}
As mentioned before, it is important to realize that particle and energy diffusion  are related phenomena.
In line with the discussion of \eqref{eqn:Q_tot},
expanding Eq.\ \eqref{eqn:qmoment} in the small velocity limit gives
\begin{align}
    q^\mu = m V_f^\mu + \frac{1}{2 m^2} \int {\rm d}^3 p \, p_{<\nu>} p^{<\nu>} p^{<\mu>} \corr{f} + \ldots.
    \label{eqn:qexpand}
\end{align}
This demonstrates that the energy diffusion 4-vector contains a diffusion contribution plus energy flux corrections orthogonal to the 4-velocity.
The latter should be compared with the heat-flux in the Newtonian version of Grad's 13-moment equations \citep{Grad13},
\begin{align}
    q_{\rm Grad}^i = \corr{\frac{m}{2}}\int {\rm d}^3 v \left(v^j - u^j\right)\left(v_j - u_j\right) \left(v^i - u^i\right) f.
    \label{eqn:qgrad}
\end{align}
In this sense, a 14-moment based closure in relativistic systems is the natural extension of Grad's 13-moment expansion.\\

In the following we will consider two limiting cases for the collision
operator. The simplest one being the absence of collisions, $C\left[ f
\right]=0$, and the strongly collisional limit described by \citet{Denicol:2012cn,Denicol:2018rbw,Denicol:2019iyh}.

\subsection{Collisional fluids}
\label{sec:col_closure}

Although the \erm{specific} choice of closure will highly depend on the properties of the
system, the general form of the closure is known in the collisional limit
\citep{Denicol:2012cn,Denicol:2018rbw,Denicol:2019iyh}.  
Starting from a perturbative description in inverse Reynolds and Knudsen numbers, ${\rm
Re}^{-1}$ and $\rm Kn$, respectively, \citet{Denicol:2012cn} introduced an
effective power counting scheme in order to expand the collision operator,
in addition to identifying (and keeping) only the most relevant time scale
associated with \erm{binary} collisions.  
More precisely, the expansion of the closure is done up to second order in 
\begin{align}
  {\rm Re}^{-1}_\Pi := \frac{\left|\Pi\right|}{P},~
  {\rm Re}^{-1}_V := \frac{\left|V_f^\mu\right|}{n_f},~
  {\rm Re}^{-1}_\pi := \frac{\left|\pi^{\mu\nu}\right|}{P},~
  {\rm Kn} := \ell_{\rm micro}/L,
  \label{eqn:Kn}
\end{align}
where $\ell_{\rm micro}$ is small scale set by microscopic time scales associated with interactions, and
$L$ is the typical system size associated with gradients of the hydrodynamic variables, such as the flow velocity.
Expressing the closure only in terms of the 14 moments, in a near-equilibrium expansion one finds that higher-rank moments
relax to higher derivatives of those moments. Hence, in this limit the higher order moments have to obey the following
scaling in terms of Knudsen numbers 
\begin{align}
    f_{-i}^{\alpha\beta\gamma\erm{\ldots}} \simeq \mathcal{O}\left( {\rm Kn}^3\right)\,,~ i>0.
    \label{eqn:fnegr_Kn}
\end{align}
This directly implies that as long as we are in the limit of high collisionality,  ${\rm Kn} \ll 1$,
neglecting all higher-rank moments with negative $r<0$ is expected to give a good approximation of the dynamics. Using a general treatment of the collisional operator
$C\left[f\right]$, up to second order in $\mathcal{O}\left({\rm Re}^{-1}\,{\rm Kn}\right)$ the closure relations read \citep{Denicol:2019iyh},
\corr{specialized here to the Eckart frame $V_f^\mu=0$, in which the energy
diffusion vector $q^\mu$ is also the heat flux. A general hydrodynamic frame
requires separate particle- and energy-diffusion terms.}
\begin{dmath}
  u^\mu \nabla_\mu {\Pi} = -\frac{1}{\tau_\Pi} \Pi - \corr{\frac{\zeta}{\tau_\Pi}} \theta - \delta_{\Pi\Pi} \Pi
  \theta + \lambda_{\Pi\pi} \pi^{\mu\nu} \sigma_{\mu\nu}
  - \ell_{\Pi q}
  \nabla_\mu q^\mu - \tau_{\Pi q} q^\mu \dot{u}_\mu - \lambda_{\Pi q} q^\mu
\nabla_\mu \alpha_0 -\delta_{\Pi q E} \mathfrak{q} q^\mu e_\mu,
  \label{eqn:PiEvol}
\end{dmath}
  \begin{dmath}
u^\alpha \nabla_\alpha{q}^{<\mu>} = -\frac{1}{\tau_q}q^\mu - \corr{\frac{\kappa}{\tau_q}} \nabla^{\erm{<\mu>}} \alpha_0
\corr{- q_\nu \omega^{\nu\mu}} + \delta_{qq} q^\mu \theta +
\lambda_{qq} q_\nu \sigma^{\mu\nu}  
+ \ell_{q\Pi} \nabla^{\erm{<\mu>}} \Pi -
\corr{\ell_{q\pi}} \Delta^{\mu\nu} \nabla_\lambda\pi^\lambda_\nu - \tau_{q\Pi} \Pi
\dot{u}^\mu + \tau_{q\pi}\pi^{\mu\nu}\dot{u}_\nu 
- \lambda_{q\Pi}\Pi
\nabla^{\erm{<\mu>}}\alpha_0 + \lambda_{q \pi} \pi^{\mu\nu}\nabla_\nu \alpha_0 +
\corr{\delta_{qB} \mathfrak{q} b^{\mu\nu}q_\nu}
- \delta_{qE} \mathfrak{q}e^\mu
- \delta_{q\Pi E} \mathfrak{q}\Pi e^\mu - \delta_{q\pi E} \mathfrak{q}
\pi^{\mu\nu}e_\nu\,,
  \label{eqn:qEvol}
  \end{dmath}
  \begin{dmath}
u^\alpha \nabla_\alpha{\pi}^{<\mu \nu>} = -\frac{1}{\tau_\pi}\erm{\pi^{\mu\nu}} \corr{- \frac{2\eta}{\tau_\pi}} \sigma^{\mu\nu}
+ 2 \pi_\lambda^{<\mu}\omega^{\nu > \lambda} - \delta_{\pi\pi} \pi^{\mu\nu} \theta - \tau_{\pi\pi} \pi^{\lambda<\mu}\sigma^{ \nu >}_\lambda
 + \lambda_{\pi\Pi} \Pi \sigma^{\mu\nu} -{\tau_{\pi q}} q^{<\mu}\dot{u}^{\nu >} +\ell_{\pi q} \nabla^{<\mu} q^{ \nu >} + \lambda_{\pi q} q^{<\mu}\nabla^{ \nu>}\alpha_0
 - \delta_{\pi B} \mathfrak{q} b^{\alpha\beta} \Delta^{\mu\nu}_{\alpha \kappa} g_{\lambda \beta}\pi^{\kappa\lambda} + \delta_{\pi q E} \mathfrak{q} e^{<\mu}q^{ \nu>}.
\label{eqn:pi_Evol}
\end{dmath}
\corr{All named second-order coefficients on the right-hand sides above are
understood to be rate-normalized by the relaxation time of their respective
equation. For example, $\delta_{\Pi\Pi}$ here denotes
$\delta_{\Pi\Pi}^{\rm(IS)}/\tau_\Pi$, and analogously for the $q$ and $\pi$
equations. The Navier--Stokes coefficients $\zeta$, $\kappa$, and $\eta$ retain
their conventional definitions, which is why their factors of $1/\tau$ are
shown explicitly.}
Here $\delta_i$, $\ell_i$, $\tau_i$ are transport coefficients that appear in this approach.
It is important to stress that while the functional form of the closure is
known, the precise values of the transport coefficients need to be
specified, as they depend on the microscopic properties of the fluid.  For the case of a massless
ultra-relativistic gas interacting via a constant cross section, they can be found in \citet{Denicol:2019iyh}.
Some of these transport coefficients have further been computed for Coulomb
collisions in Newtonian plasmas (e.g., \citealt{kulsrud2020plasma}).

\subsection{Collisionless fluids}

Although introducing the 14-moment decomposition might at first glance
imply that the system may be in a hydrodynamic regime, it is
important to understand that \erm{if effective collisions are only mediated by the collectively created mean electromagnetic fields} an equilibrium might be reached but its form is not known a priori. As a direct consequence, the non-perfect fluid equivalents, $V_f^\mu$, $q^\mu$ and
$\pi^{\mu\nu}$, can not only be large, but there is not necessarily a well-defined isotropic diagonal
pressure component $P=P\left( e,n_f,\dots \right)$. 
Instead, $f_0^{\mu\nu}$ contains all components of
the pressure with off-diagonal elements that can be as large or even exceed the
diagonal ones.\\

It is trivial to show that Eq.\ \eqref{eqn:BoltzmannR} implies conservation
of particle number and energy-momentum, i.e.
\begin{align}
&  \nabla_\mu \left( n_f u^\mu + V_f^\mu \right) \,=\,0\,, \\
&  \nabla_\mu \left( e \erm{u^{\mu}u^\nu} + \left( P + \Pi \right) \Delta^{\mu\nu}
  +\erm{q^\mu u^\nu} + \erm{q^\nu u^\mu} + \pi^{\mu\nu}\right) \,=\,\corr{+ \mathfrak{q} F^{\nu\lambda}N_\lambda}\,,
  \label{eqn:EMF_n}
\end{align}
The missing evolution equations for the diffusion current and the stresses
have to be derived from the Boltzmann equation \eqref{eqn:BoltzmannR}.

Assuming that the rest-mass of the particles is larger than any other scales, which includes the \emph{effective} temperature of the plasma in the nearly-collisionless limit, one may consider $p_{<\mu>} p^{<\mu>} \ll m^2$ and Taylor-expand the square-root in Eq.\
\eqref{eqn:f2}, so as to obtain the following recursion relation
\begin{align}
  f^{\mu_1\dots\mu_n}_{r+s} = m^s \left( f^{\mu_1\dots\mu_n}_{r} + \frac{s}{2 m^2}
f_{r}^{\alpha\beta
  \mu_1\dots\mu_n}\Delta_{\alpha\beta} + \mathcal{O}\left( \frac{1}{m^4}\right)
  \right).
  \label{eqn:frecursive}
\end{align}
It is important to understand that this assumption, which is akin to considering  non-relativistic effective temperatures
in the r-closure relation \eqref{eqn:frecursive}, does not break the covariance
of the moment equations. \corr{For a non-degenerate gas the expansion parameter
is $\langle p_{\langle\mu\rangle}p^{\langle\mu\rangle}\rangle/m^2\sim3T/m$,
so the closure relations below have relative errors of order $T/m$ and are
controlled only for $T/m\ll1$. The closed system therefore describes
non-relativistic species temperatures but allows arbitrarily relativistic bulk
flow. Applied per species, it requires $T_e/m_e\ll1$ and $T_p/m_p\ll1$
separately and does not describe relativistically hot electrons or pair
plasmas. The exact master hierarchy derived below remains valid outside this
closure regime, but must then be closed differently; see also
\citet{Tinti:2018qfb}.}

In order to obtain the moment reduction, we note that
\begin{align}
  p^\mu \partial_\mu f = - \left( p_\alpha u^\alpha \right) u^\mu \nabla_\mu f
  + p^{\left<\mu\right>} \nabla_\mu f\,,
  \label{eqn:pf}
\end{align}
implies that an irreducible moment decomposition will naturally lead
to an Israel-Stewart-type equation \citep{Israel:1979wp}, i.e. equations of motion of relaxation-type containing $u^\mu \nabla_\mu
\pi^{\alpha\beta}$, $u^\mu \nabla_\mu \Pi$, $u^\mu \nabla_\mu V_f^\alpha$, and $u^\mu \nabla_\mu q^\alpha$ terms.
Since the total heat flux combines particle diffusion and energy diffusion,
it is only natural to consider the evolution of 
\begin{align}
    \mathcal{G}^\mu = q^\mu - m V_f^\mu,
\end{align}
for a collisionless plasma (this is also in line with \citealt{Denicol:2019iyh}).

Following \citet{Tinti:2018qfb}, taking the corresponding irreducible moments of \eqref{eqn:BoltzmannR} with
$C[f]=0$ one obtains the general master equations
\begin{align}
\corr{\Delta^{\mu}_{\nu} u^\alpha \nabla_\alpha f_r^{\nu}} &= \corr{- \theta f_r^\mu
- \Delta^\mu_\nu \nabla_\alpha f_{r-1}^{\nu\alpha}
+ \mathfrak{q} f_r\, e^\mu} \nonumber\\
&\ \ \corr{+ \mathfrak{q} b^{\mu\alpha} g_{\alpha\beta} f_{r-1}^{\beta}
+ \mathfrak{q}\left(r-1\right) e_\alpha f_{r-2}^{\mu\alpha}} \nonumber\\
&\ \ \corr{- \left(r-1\right) \dot{u}_\alpha f_{r-1}^{\mu\alpha}
- \left(r-1\right) \nabla_\alpha u_\beta f_{r-2}^{\mu\alpha\beta}} \nonumber\\
&\ \ \corr{- f_{r+1}\, \dot{u}^\mu
- \Delta^\mu_\nu \left(\nabla_\alpha u^\nu\right) f_r^{\alpha}}\,,
\label{eqn:master_rank1}
\end{align}
\begin{align}
\corr{\Delta^{\mu}_{\alpha}\Delta^{\nu}_{\beta} u^\lambda \nabla_\lambda f_r^{\alpha\beta}} &= \corr{- \theta f_r^{\mu\nu}
- \Delta^\mu_\alpha \Delta^\nu_\beta \nabla_\lambda f_{r-1}^{\alpha\beta\lambda}} \nonumber\\
&\ \ \corr{+ 2\mathfrak{q}\, e^{\left(\mu\right.} f_r^{\left.\nu\right)}
+ 2 \mathfrak{q} b^{\left(\mu\right.}{}_{\lambda} f_{r-1}^{\left.\nu\right)\lambda}} \nonumber\\
&\ \ \corr{+ \mathfrak{q}\left(r-1\right) e_\lambda f_{r-2}^{\mu\nu\lambda}
- \left(r-1\right) \dot{u}_\lambda f_{r-1}^{\lambda\mu\nu}} \nonumber\\
&\ \ \corr{- \left(r-1\right) \nabla_\lambda u_\rho f_{r-2}^{\lambda\rho\mu\nu}
- 2 \dot{u}^{\left(\mu\right.} f_{r+1}^{\left.\nu\right)}} \nonumber\\
&\ \ \corr{- 2 \left(\nabla_\lambda u^{\left(\mu\right.}\right) f_r^{\left.\nu\right)\lambda}}\,,
\label{eqn:master_rank2}
\end{align}

\corr{The evolution of $\mathcal{G}^\mu$ follows from \eqref{eqn:master_rank1}
for $r=1$ minus $m$ times $r=0$, and the evolution of $\Pi^{\mu\nu}=f_0^{\mu\nu}$ from
\eqref{eqn:master_rank2} with $r=0$,}
  \begin{dmath}
u^\alpha \nabla_\alpha \mathcal{G}^{<\mu>} = \corr{+} \corr{\mathfrak{q}} e^{\mu} \left( f_1 - m f_0\right)
\corr{+ m}\, \mathfrak{q} e_\alpha f_{-2}^{\alpha\mu} + \mathfrak{q}
b^{\mu\alpha}g_{\alpha\beta} \left( V_f^\beta - m f_{-1}^\beta \right) - \theta \mathcal{G}^{<\mu>}
\corr{- m \dot{u}_\alpha f_{-1}^{\alpha \mu}}
\corr{- \left( e - nm\right)\dot{u}^\mu} - \Delta^\mu_\nu \nabla_\alpha \left( \Pi^{\nu \alpha} - m f_{-1}^{\nu\alpha} \right) - \Delta^\mu_\nu \left(\nabla_\alpha u^\nu\right) \mathcal{G}^{\alpha}
\corr{- m}\,
\nabla_\alpha u_\beta f_{-2}^{\alpha\beta\mu}\,,
\label{eqn:G_exact}
\end{dmath}
\begin{dmath}
\corr{\Delta^\mu_\rho\Delta^\nu_\sigma u^\alpha \nabla_\alpha \Pi^{\rho\sigma}} = \corr{+ 2 \mathfrak{q}} e^{\left(\mu\right.}V_f^{\left.\nu\right)}
\corr{- \mathfrak{q}} e_\alpha f_{-2}^{\alpha\mu\nu} + 2 \corr{\mathfrak{q}}
b^{\left(\mu\right.\alpha}f_{-1}^{\left.\nu\right)\beta}g_{\alpha\beta} - \theta \Pi^{\mu\nu}
-2 u^\alpha\nabla_\alpha u^{\left(\mu\right.}f_{1}^{\left.\nu\right)} -
\nabla_\alpha f_{-1}^{\alpha <\mu><\nu>} 
\corr{+ \dot{u}_\alpha f_{-1}^{\alpha\mu\nu}}
-2 \nabla_\alpha u^{\left(\mu\right.} \Pi^{\left.\nu\right)\alpha} \corr{+}
\nabla_\alpha u_\beta f_{-2}^{\alpha\beta\mu\nu}\,.
\label{eqn:Pi_exact}
\end{dmath}
 We have also used that $\dot{u}^\alpha = u^\mu \nabla_\mu u^\alpha$.
We can see that the evolution of the hydrodynamic moments depends on higher
order moments in $r<0$. Using our proposed closure relation valid in the \erm{\corr{low-temperature regime}}, Eq.\ \corr{\eqref{eqn:frecursive}}, we can re-express them as
follows. \corr{Applying \eqref{eqn:frecursive} to the rank-one and rank-two moments,
together with the exact contraction $\Delta_{\alpha\beta}f_{r}^{\alpha\beta\mu_1\dots}=f_{r+2}^{\mu_1\dots}-m^2 f_{r}^{\mu_1\dots}$,
gives the closure relations}
\begin{align}
  \corr{m f_{-1}^{\mu}} &\corr{{}\simeq 2 V_f^\mu - \frac{1}{m} q^\mu\,,\quad
  m f_{-1}^{\mu\nu} \simeq m^2 f_{-2}^{\mu\nu} \simeq \Pi^{\mu\nu}}\,, \nonumber\\
  \corr{m\left(f_1 - m f_0\right)} &\corr{{}\simeq e - m n \simeq \frac{3}{2}\left(P+\Pi\right)}\,,
  \label{eqn:closure_rels}
\end{align}
\corr{while the independent symmetric trace-free parts of moments of rank $\geq 3$
with $r<0$ are dropped according to \eqref{eqn:fnegr_Kn}; their traces still obey
the exact on-shell contraction above. Note that
$V_f^\mu - m f_{-1}^\mu = \mathcal{G}^\mu/m$, i.e.\ the
magnetic term in \eqref{eqn:G_exact} changes sign relative to the naive substitution
$m f_{-1}^\mu \to q^\mu/m$. With this closure one finds}
  \begin{dmath}
u^\alpha \nabla_\alpha \mathcal{G}^{<\mu>} = \corr{+} \frac{\mathfrak{q}}{m} e^{\mu} \left( e -m n\right)
\corr{+ \frac{\mathfrak{q}}{m} e_\alpha \Pi^{\alpha\mu}} \corr{+ \frac{\mathfrak{q}}{m}
b^{\mu\alpha}\mathcal{G}_{\alpha}} - \theta \mathcal{G}^{\mu}
\corr{- \dot{u}_\alpha \Pi^{\alpha \mu}} \corr{- \left(e-mn\right)\dot{u}^\mu}
 \corr{- \frac{1}{2 m^2}\Delta^\mu_\nu \nabla_\alpha f_{0\,\lambda}^{\lambda\nu\alpha}} - \Delta^\mu_\nu \left(\nabla_\alpha u^{\erm{\nu}}\right) \mathcal{G}^{\erm{\alpha}}\,,
\label{eqn:V_less}
\end{dmath}
\begin{dmath}
\corr{\Delta^\mu_\rho\Delta^\nu_\sigma u^\alpha \nabla_\alpha \Pi^{\rho\sigma}} = \corr{+ 2 \mathfrak{q}} e^{\left(\mu\right.}V_f^{\left.\nu\right)}
+ 2 \frac{\corr{\mathfrak{q}}}{m}
b^{\left(\mu\right.\alpha}\Pi^{\left.\nu\right)\beta}g_{\alpha\beta} - \theta \Pi^{\mu\nu}
\corr{-2 \dot{u}^{\left(\mu\right.} q^{\left.\nu\right)}}
-2 \nabla_\alpha u^{\left(\mu\right.} \Pi^{\left.\nu\right)\alpha}
\corr{- \frac{1}{m}\Delta^\mu_\alpha\Delta^\nu_\beta \nabla_\lambda \left(\Pi^{\alpha\beta\lambda}_{(-1)}\right)}\,,
\label{eqn:Pi_less}
\end{dmath}

In accordance with our previous discussion, we can further split the last
equation into its trace and trace-free part. This then reads,

\begin{dmath}
  u^\alpha \nabla_\alpha \corr{\left(P+\Pi\right)} = \corr{+ \frac{2}{3} \mathfrak{q}\, e_\mu V_f^\mu}
  \corr{-  \frac{5}{3}\theta \left(P+\Pi\right)}
  \corr{-\frac{2}{3} q^\mu \dot{u}_\mu}
  \corr{-\frac{2}{3} \nabla_\mu \mathcal{G}^\mu}
-\frac{2}{3} \sigma_{\mu\nu} \pi^{\mu\nu} \,.
\label{eqn:big_pi_less}
\end{dmath}

\begin{dmath}
  u^\alpha \nabla_\alpha \pi^{<\mu\nu>} = \corr{+ 2 \mathfrak{q}}\,
  e^{<\mu} V_f^{\nu>}
  \corr{+ 2 \frac{\mathfrak{q}}{m}
 b^{<\mu}_{\ \ \alpha}\pi^{\nu>\alpha}}
 - \theta \pi^{\mu\nu} 
\corr{-2  \dot{u}^{<\mu} q^{\nu>}}
-2 \Delta^{\mu\nu}_{\gamma\delta} \nabla_\alpha u^{\gamma} \pi^{\delta \alpha} 
\corr{- 2\left(P+\Pi\right) \sigma^{\mu\nu}}\,.
\label{eqn:small_pi_less}
\end{dmath}

While these equations are fully generic and self-consistent, they do require the specification of additional closure relations for the higher than second moments, with negative values for $r$.
We will provide one such a potential closure in Sec. \ref{sec:coll_less}.

\subsection{Local relaxation closure for collisionless fluids}
\label{sec:coll_less}

As a particular application of a collisional closure to a collisionless
fluid, we follow the approach of \citet{2015PhPl...22a2108W} and adopt an isotropic
pressure closure. The idea is to damp anisotropic pressure contributions
over a damping time $\tau_\pi$. Since we evolve $\Pi$ and $\pi^{\mu\nu}$ separately,
this only results in the addition of a damping term to Eq. \eqref{eqn:small_pi_less}.
Fully consistent with the application of a collisional closure in the Newtonian approach, we further
apply the additional collisional assumption that
higher-rank moments can be neglected according to Eq. \eqref{eqn:fnegr_Kn}. 
Although this assumption might appear ad-hoc at first, in particular when the Knudsen number
becomes comparable to unity, comparisons with particle-in-cell simulations of the 
Vlasov equation for Newtonian plasmas have shown reasonable agreement with such a closure approach 
in a 10-moment formulation \citep{2015PhPl...22a2108W}. For convenience, we will also introduce collisional
damping times for the particle diffusion and the energy diffusion current.
Furthermore, we will work within the Eckart frame 
and drop the particle diffusion current $V_f^\mu$ in favor of the heat flux vector $q^\mu$. 
We then find that expressions
\eqref{eqn:V_less}, \eqref{eqn:big_pi_less} and \eqref{eqn:small_pi_less}
reduce to
 \begin{dmath}
   u^\alpha \nabla_\alpha q^{<\mu>} = \corr{+} \frac{\mathfrak{q}}{m} \left(e -m n  + P+ \Pi \right) e^{\mu}
   \corr{+} \frac{\mathfrak{q}}{m} e_\alpha \pi^{\alpha\mu}
   \corr{+} \frac{\mathfrak{q}}{m} b^{\mu\alpha}q_\alpha
- \left(e-m n + P+\Pi\right) \dot{u}^{\mu}
- \theta q^{\mu} 
 \corr{-} \dot{u}_\alpha \pi^{\mu\alpha} - q^{\alpha}\nabla_\alpha u^\mu -\frac{1}{\tau_q} q^\mu\,,
\label{eqn:q_less_col}
\end{dmath}

\begin{dmath}
  u^\alpha \nabla_\alpha \left(P+\Pi\right) =
  \corr{- \frac{2}{3}} \nabla_\mu q^\mu
  -  \frac{\corr{5}}{3}\theta \left(P+\Pi\right)
  -\frac{2}{3} q^\mu \dot{u}_\mu
-\frac{2}{3} \sigma_{\mu\nu} \pi^{\mu\nu} \,,
\label{eqn:big_pi_less_col}
\end{dmath}

\begin{dmath}
  u^\alpha \nabla_\alpha \pi^{<\mu\nu>} =
   + 2 \frac{\corr{\mathfrak{q}}}{m}
 \corr{b^{<\mu}_{\ \ \alpha}\pi^{\nu>\alpha}}  - \theta \pi^{\mu\nu}
-2 \dot{u}^{<\mu} q^{\nu>}
-2  \nabla_\alpha u^{<\mu} \pi^{\nu> \alpha} 
\corr{- 2}\left(P+\Pi\right) \sigma^{\corr{\mu\nu}}
-\frac{1}{\tau_\pi} \pi^{\mu\nu}\,.
\label{eqn:small_pi_less_col}
\end{dmath}

Comparing the above expressions with the generic collisional closure
discussed in Sec.\ \ref{sec:col_closure}, we can see that \corr{within the
low-temperature truncation} the collisionless
system becomes a particular \erm{variant} of a collisional system, consistent
with the assumption of neglecting higher order moments and introducing
relaxation times. {Most importantly, the collisionless moment expansion
together with the higher-moment truncation fixes the coefficients of the closure exactly}. {We note that while other closure relations are in use in the Newtonian plasma community (e.g., \citealt{PhysRevLett.64.3019}) their non-local nature renders them unsuitable for use with relativistic approaches.}
Thus, for such an isotropic closure we can
easily use the same tools that will be developed in the rest of this paper,
allowing us to describe a unified system of equations for two-fluid
collisional plasmas and \corr{low-temperature collisionless closures}.

\section{Relativistic two-fluid magnetohydrodynamics}
\label{sec:twofluids}

{Having discussed how to model out-of-equilibrium single component fluids, we now want to recast those equations in a form suitable to describe multi-fluid plasmas.} 
More precisely, we aim to describe a two-component plasma consisting of
electrons $(e)$ and ions $(p)$, with charge number $Z$. We use
$\mathfrak{q}_e>0$ for the elementary charge magnitude and
$\corr{\overline{\mathfrak{q}}_e=-\mathfrak{q}_e}$,
$\corr{\overline{\mathfrak{q}}_p=+Z\mathfrak{q}_e}$ for the signed species
charges.
Here $\mathfrak{q}_e$ refers to the electron and $\mathfrak{q}_p$ to the ion characteristic charge, respectively.
{While we adopt the convention of referring to the second species as
\textit{ions}, all results would equally be valid also for positrons.}
Instead of evolving each component separately, it will turn out to be most
beneficial to introduce a single fluid reference frame. This will allow us
to replace one of the species with an effective single fluid for which
total energy momentum conservation, together with the conservation of the
electromagnetic sector can be enforced. Such single fluid descriptions have been
discussed, e.g., by \citet{Koide:2009yx,Andersson:2021kfk}, and \citet{cercignani2002relativistic}, 
whose notation we follow. \\

We start out by providing a brief description of both fluids using the
language of transient (magneto-) hydrodynamics.
Each species, electrons and ions, will have separate particle four-currents
{and energy-momentum tensors}, viz.
\begin{align}
    & N_e^\mu\,,\ T_e^{\mu\nu}\ {\rm (electrons)}\,, \label{eqn:defNe}\\
    & N_p^\mu\,,\ T_p^{\mu\nu}\ {\rm (ions/positrons)}\,. \label{eqn:defNp}
\end{align}
Within this description, each species has their own
rest-frame associated with $N^\mu_{e/p}$. Within the description of a
single fluid, it turns out to be more useful to introduce a mass-weighted
average velocity to be used as the rest-frame of the joint single fluid.
This corresponds to the center of mass frame in which collisions between the
two species take place.
We define this via
\begin{align}
  N^\mu = \chi_e N_e^\mu + \chi_p N_p^\mu =: n U^\mu,
  \label{eqn:Njoint}
\end{align}
where the joint number density $n$ is defined by demanding that $U^\mu U_\mu
=-1$. The coefficients can be chosen freely depending on the particular
fluid and equilibrium being described. Example choices include averaging
$\chi_e = \chi_p=1$, or mass-weighted averages $\left( \chi_e = m_e/\left(
m_e + m_p \right)\,,
\chi_p = m_p/\left( m_e+m_p \right)\right)$, where $m_e$ and $m_p$ are the
electron and ion masses, respectively.
Adopting $U^\mu$ as the primary velocity frame, we can recast
\eqref{eqn:defNe} and \eqref{eqn:defNp} into
\begin{align}
N_e^\mu = n_e U^\mu + V_e^\mu,
  \label{eqn:N2e}\\
N_p^\mu = n_p U^\mu + V_p^\mu.
  \label{eqn:N2p}
\end{align}
As a consequence of using a joint reference frame that does not align with
the comoving frame of either species, we can see that now both species
develop diffusion currents $V^\mu_{e/p}$.
By means of \eqref{eqn:Njoint}, these will not be independent but are
related via
\begin{align}
  V_p^\mu = - \frac{\chi_e}{\chi_p}V_e^\mu.
  \label{eqn:diffrel}
\end{align}
Furthermore, we may split the electric 4-current into its individual
contributions
\begin{align}
  \mathcal{J}^\mu &= \mathfrak{q}_e \left( Z N_p^\mu - N_e^\mu \right)\,, \nonumber\\
  &= \mathfrak{q}_e\left[ \left( Z n_p - n_e \right) U^\mu - \left( 1 + Z
  \frac{\chi_e}{\chi_p} \right) V_e^\mu\right]\,,
  \label{eqn:Jsplit}
\end{align}
where $\mathfrak{q}_e$ denotes the electric charge. We have further made use of
\eqref{eqn:diffrel} when re-expressing this equation.

Each species, can be described my means of their energy momentum tensors,
$T^{\mu\nu}_e$ and $T^{\mu\nu}_p$.
Decomposing each of these using the common fluid velocity $U^\mu$, these
are given by \citep{cercignani2002relativistic}
\begin{align}
  T^{\mu\nu}_X \,=&\, e_X U^\mu U^\nu + U^\mu \left( q_X^\nu + h_X V^\nu_X\right)
  +U^\nu \left( q_X^\mu + h_X V^\mu_X\right)\nonumber  \\
&  + \left( p_X + \Pi_X \right)
  \Delta^{\mu\nu} + \pi_X^{\mu\nu}.
  \label{eqn:Tindiv}
\end{align}
Here, $X=e\,,p$ refers to the individual species.
This decomposition differs from \eqref{eqn:Tfsplit} in several important
ways. First, the constraints \eqref{eqn:pi} apply with respect to $U^\mu$
for each species $X$. Second, because we have adopted a frame that does not
align with the individual Eckart frame of each species, an additional
contribution to the heat flux, $h_X V_X^\mu$, appears, corresponding to the
diffusion of that species in the frame described by $U^\mu$, see also the discussion
around Eq. \eqref{eqn:Q_tot}.

The combined single-fluid energy momentum tensor, can then be expressed as
\begin{align}
  T_f^{\mu\nu} = T^{\mu\nu}_e + T^{\mu\nu}_p.
  \label{eqn:tsum}
\end{align}
Using the decomposition lined out in \eqref{eqn:Tfsplit},
\begin{align}
  T_f^{\mu\nu} = e U^\mu U^\nu + \left( P+\Pi \right) \Delta^{\mu\nu} +
  q^\mu U^\nu + q^\nu U^\mu + \pi^{\mu\nu}\,,
  \label{eqn:Tfsplit_repeat}
\end{align}
we can easily
identify,
\begin{align}
&  e = e_e + e_p, \label{eqn:esum}\\
&  p = p_e + p_p, \label{eqn:psum}\\
&  \Pi = \Pi_e + \Pi_p, \label{eqn:pisum}\\
&  q^\mu = q^\mu_e + q^\mu_p + h_e V_e^\mu + h_p V_p^\mu =  q^\mu_e +
q^\mu_p + \left( h_e - h_p\frac{\chi_e}{\chi_p} \right) V_e^{\corr{\mu}}, \label{eqn:qsum}\\
&  \pi^{\mu\nu} = \pi_e^{\mu\nu} + \pi_p^{\mu\nu}. \label{eqn:ppisum}
\end{align}
We provide a more detailed discussion of how the single fluid frame relates to the component frames in Appendix \ref{app:transform}.
\noindent Overall, conservation of total energy-momentum \eqref{eqn:T_total} then
implies,
\begin{align}
  \nabla_\mu \left( T_f^{\mu\nu} + T_{\rm EM}^{\mu\nu} \right)=0.
  \label{eqn:Tmunutot}
\end{align}

We point out when solving these equations no electromagnetic source terms
are present, and energy and momentum are exactly conserved.
As such, this description is really that of a single fluid coupled to
electromagnetism. However, the two fluid nature of the system is
fundamentally encoded in the heat fluxes and anisotropic stresses present in the system.

%

\subsection{Two-fluid interactions and electron diffusion}

Following the discussion of how to combine a relativistic two-fluid plasma
into a single fluid description, we are left to specify evolution equations
for the out-of-equilibrium variables. In particular, we will need to specify
internal heat fluxes $q_X^\mu$ of the plasma, the bulk viscous scalar
$\Pi$ and the anisotropic stresses $\pi^{\mu\nu}_X$.
In addition, we strictly require prescriptions of the electron diffusion
current $V_e^\mu$ and the electron enthalpy $h_e$, as these are required in
order to determine the total heat flux $q^\mu$, see Eq. \eqref{eqn:qsum}.
Different from the other variables, it is straightforward to determine
evolution equations for $V_e^\mu$ and $h_e$.

When writing the electron evolution equation, we want to account for potential electron-ion collisions. {We denote such a term as $\mathcal{C}_e^\nu$.
In the absence of explicit collisions $\mathcal{C}_e^\nu = 0$.}

Overall, the evolution equations for the electron \erm{read}
\begin{align}
  \nabla_\mu T^{\mu\nu}_e + \mathfrak{q}_e F^{\corr{\nu\mu}} N_{e\,\mu} = \mathcal{C}_e^\nu.
  \label{eqn:collisions}
\end{align}

We can alternatively write expression \eqref{eqn:collisions} as 
\begin{align}
  \nabla_\mu T^{\mu\nu}_e =& - \mathfrak{q}_e n_e e^\nu \corr{+}\mathfrak{q}_e
  b^{\corr{\mu\nu}} V_{e\,\corr{\mu}} - \mathfrak{q}_e U^\nu V_e^\mu e_\mu
     + \mathcal{C}_e^\nu.
  \label{eqn:dT_e2}
\end{align}

These equations describe the evolution of the electron fluid. However,
using the decomposition with respect to the single fluid frame, see Eq.
\eqref{eqn:Tindiv} for $X=e$, it does not describe the evolution of the
single fluid frame velocity $U^\mu$ or energy density $e$, as would be the
case for a single fluid. Instead, the conservation of electron energy and
momentum determines the evolution of the electron diffusion current
$V_e^\mu$ and the the electron enthalpy $h_e$ in the single-fluid frame.

We can make this more explicit by splitting \eqref{eqn:dT_e2} along the
joint fluid velocity $U^\mu$.
More specifically, by contracting \eqref{eqn:dT_e2} with $U_\nu$, we find that
\begin{align}
  U^\mu \nabla_\mu e_e =& - \left( e_e + P_e + \Pi_e\right) \nabla_\mu U^\mu
  \corr{-} \pi_e^{\mu\nu}\sigma_{\mu\nu} \corr{-} \nabla_\mu \mathcal{Q}_e^\mu
  \nonumber \\
  &- \mathfrak{q}_e e_\mu
V_e^\mu \corr{-} \mathcal{Q}_e^\nu U^\mu \nabla_\mu U_\nu \corr{- U_\nu \mathcal{C}_e^\nu},
  \label{eqn:electron_energy}
\end{align}
where we have introduced the shear tensor
$\sigma_{\mu\nu}=
\Delta_{\mu\nu}^{\alpha\beta} \nabla_\alpha
U_\beta$.
At the same time, we can obtain an evolution equation for $V_e^\mu$, by
contracting \eqref{eqn:dT_e2} with $\Delta^\mu_{\ \nu}$,
\begin{align}
  -\Delta^\mu_\nu U^\lambda \nabla_\lambda \mathcal{Q}_e^\nu =&  \left( e_e + P_e
  + \Pi_e \right) U^\lambda \nabla_\lambda U^\mu + \Delta^\mu_\nu
  \nabla_\lambda \pi_e^{\lambda\nu}\nonumber\\
  &\corr{+ \mathcal{Q}_{e\, \nu}
  \sigma^{\mu\nu} - \mathcal{Q}_{e\, \nu} \omega^{\mu\nu} + \frac{4}{3}}
\mathcal{Q}_e^\mu \nabla_\nu U^\nu  \nonumber\\ 
  &+\Delta^{\mu\nu} \nabla_\nu \left( P_e
+\Pi_e \right) + \mathfrak{q}_e n_e e^\mu \corr{+} \mathfrak{q}_e
  b^{\mu\nu}V_{e\, \nu} \nonumber\\
  & \corr{- \Delta^\mu_\nu \mathcal{C}_e^\nu}
  \label{eqn:electron_momentum}
\end{align}
where we have introduced the vorticity tensor $\omega_{\mu\nu}= \frac{1}{2}
\corr{\Delta_{\mu}^{\alpha}\Delta_{\nu}^{\beta}} \left( \nabla_\alpha U_\beta - \nabla_\beta
U_\alpha\right)$ and the total electron \erm{energy diffusion vector} 
\begin{align}
\mathcal{Q}_e^\mu = q_e^\mu + h_e V_e^\mu.
\end{align}

It is important to stress at this point that Eqs.\ \eqref{eqn:electron_energy} and \eqref{eqn:electron_momentum}
have exactly the same form as dissipative bulk pressures \eqref{eqn:PiEvol} and energy diffusion fluxes \eqref{eqn:qEvol}.
As such, the electron fluid contributions appear as \erm{14-moment} dissipative corrections to the effective single fluid.
This fundamentally implies that relativistic two fluid system can be handled in exactly the same way as dissipative single fluids \citep{Most:2021oha}.
\erm{Moreover, expression \eqref{eqn:electron_momentum} is nothing but the general Ohm's law we set out to derive in \eqref{eqn:Ohm}.
Compared to a Newtonian Ohm's law, the main difference is that \eqref{eqn:electron_momentum} provides a time evolution equation for the
dissipative electric current $V_e^\mu$.}
This completes the description of a two-fluid plasma within the framework
of a single fluid with dissipative corrections.
Together with the single fluid evolution equation \eqref{eqn:Tmunutot},
the electron fluid equations \eqref{eqn:electron_energy} and  \eqref{eqn:electron_momentum}
provide evolution equations for the diffusive heat flux and electron
energy.

\subsection{Electron-ion collisions}

To model inter-species collisions, we proceed as follows.
Instead of parameterizing them directly, we adopt an effective relaxation time approach, meaning that collisions will drive the electron energy-momentum tensor $T^{\mu\nu}_e$ towards its equilibrium value $T^{\mu\nu}_{e\,,\rm eq}$, see Eq. \eqref{eqn:Teq}, on a time scale $\tau$.
Following \citet{cercignani2002relativistic}, we can write the collision term as follows,

\begin{align}
   \mathcal{C}_e^\nu =  \corr{+} \frac{1}{\tau}
  \left( T^{\mu\nu}_e - T^{\mu\nu}_{e,eq} \right) U_{ \mu}\,.
  \label{eqn:CRTA}
\end{align}

Since the equilibrium solution $T^{\mu\nu}_{e,eq}$ is defined in the
absence of heat fluxes, and the energy density $e_e$ is the same in the
given frame, the collision term is given by the heat flux only.
We can alternatively write expression \eqref{eqn:CRTA} as \citep{cercignani2002relativistic}
\begin{align}
 \mathcal{C}_e^\nu =
    \corr{-\frac{1}{\tau}\mathcal{Q}_e^\nu
    =-\frac{1}{\tau}q_e^\nu-\frac{h_e}{\tau}V_e^\nu},
  \label{eqn:dT_C}
\end{align}
We impose $\corr{\mathcal{C}_p^\nu=-\mathcal{C}_e^\nu}$ so that inter-species
collisions conserve the total energy--momentum. The form in
\eqref{eqn:dT_C} follows directly when $T^{\mu\nu}_{e,\rm eq}$ is constructed
with the joint four-velocity $U^\mu$ used in \eqref{eqn:CRTA}.
Further work is needed when modeling this effective relaxation time approach when considering the case where the relaxation scale $\tau$ is momentum dependent. In this case, a modified version of the collision term in the relaxation time approximation has to be used to ensure agreement with general properties of the Boltzmann equation, see \citet{Rocha:2021zcw}.

\subsection{Summary}
{
After discussing how to recast two relativistic fluids into a single fluid form with potentially large out-of-equilibrium corrections, we want to briefly summarize the main equations as follows.
}
\erm{The electromagnetic fields are evolved using the Maxwell equation coupled to the current provided by both species,}
{
\begin{dmath}
\nabla_\mu F^{\nu\mu} = 4\pi \mathfrak{q}_e \left[ \left(Z n_p -n_e\right) U^{\corr{\nu}} - \left(1+ Z \frac{\chi_e}{\chi_p}\right) V_e^{\corr{\nu}}\right]\,.
\end{dmath}
\erm{Total conservation of energy and momentum then implies that }
\begin{dmath}
\nabla_\mu \left( \left[e_e +e_p\right] U^\mu U^\nu + \left( P_e + P_p +\Pi_e + \Pi_p \right) \Delta^{\mu\nu} +
  \left[\mathcal{Q}_e^\mu + q_p^\mu - \frac{\chi_e}{\chi_p} h_p V_e^\mu\right] U^\nu + \left[\mathcal{Q}_e^\nu + q_p^\nu - \frac{\chi_e}{\chi_p} h_p V_e^\nu\right] U^\mu + \pi^{\mu\nu} + T_{\rm EM}^{\mu\nu}\right) =0\,.
\end{dmath}
}
\erm{It is important to stress that the single fluid energy-momentum and the electromagnetic fields are conserved together.
This equation, thus has the character of a single fluid coupled to electromagnetism, where two-fluid corrections appear as dissipative heat fluxes and pressures. These obey a 14-moment like evolution equation consistent with the collisional closure outlined in Sec. \ref{sec:col_closure}. More specifically, in analogy with a bulk scalar pressure, the evolution electron energy $e_e$ is governed by}
{
\begin{align}
  U^\mu \nabla_\mu e_e =& - \left( e_e + P_e + \Pi_e\right) \nabla_\mu U^\mu
  \corr{-} \pi_e^{\mu\nu}\sigma_{\mu\nu} \corr{-} \nabla_\mu \mathcal{Q}_e^\mu
  \nonumber \\
  &- \mathfrak{q}_e e_\mu
V_e^\mu \corr{-} \mathcal{Q}_e^\nu U^\mu \nabla_\mu U_\nu \corr{- U_\nu\mathcal{C}_e^\nu}\,.
  \label{eqn:electron_energy2}
\end{align}
}
\erm{Similarly, the electron heat fluxes take the following form}
{
\begin{align}
  -\Delta^\mu_\nu U^\lambda \nabla_\lambda \mathcal{Q}_e^\nu =&  \left( e_e + P_e
  + \Pi_e \right) U^\lambda \nabla_\lambda U^\mu + \Delta^\mu_\nu
  \nabla_\lambda \pi_e^{\lambda\nu}\nonumber\\
  &\corr{+ \mathcal{Q}_{e\, \nu}
  \sigma^{\mu\nu} - \mathcal{Q}_{e\, \nu} \omega^{\mu\nu} + \frac{4}{3}}
\mathcal{Q}_e^\mu \nabla_\nu U^\nu  \nonumber\\ 
  &+\Delta^{\mu\nu} \nabla_\nu \left( P_e
+\Pi_e \right) + \mathfrak{q}_e n_e e^\mu \corr{+} \mathfrak{q}_e
  b^{\mu\nu}V_{e\, \nu} \nonumber\\
  & \corr{- \Delta^\mu_\nu \mathcal{C}_e^\nu}
  \label{eqn:electron_momentum2}
\end{align}
}{
Similar to their non-relativistic equivalent, a choice of closure \erm{for the individual species} is still
needed. That is, constitutive relations for the anisotropic stresses,
($\pi_e^{\mu\nu}, \pi_p^{\mu\nu}$), heat fluxes ($q_e^\mu,q_p^\nu$) and
pressures $\left( P_e,P_p, \Pi_e, \Pi_p \right)$ need to be provided.
\erm{These are different from dissipative corrections arising purely by the presence of a second fluid and its interaction
through the electromagnetic field.}
}
\section{Two-Fluid magnetohydrodynamics for relativistic electron-ion plasmas}
\label{sec:eI}
Having recast the two fluid equations into an effective single fluid
description with (potentially large) dissipative corrections, we want to
apply the formalism to an electron-ion plasma. Such a plasma is characterized by
the ion mass greatly exceeding the electron mass, $m_p \gg m_e$.

We begin by choosing $\chi_p = m_p/\erm{\bar{m}}$ and $\chi_e = m_e/\erm{\bar{m}}$, with
$\bar{m}=\left( m_e+m_p \right)$. This mass weighted averaging is a common choice
in Newtonian plasma physics (e.g., \citealt{sturrock1994plasma}). In practical terms, since $m_p \gg m_e$, the single-fluid frame almost aligns with the ion-frame, where the difference is given by electron contributions. Hence the dissipative correction of the presence of electrons onto the ions is small, since the effective inverse Reynolds number associated with out-of-frame corrections will be of order $m_e/m_p \ll 1$.\\
The effective particle number then is
\begin{align}
  N^\mu &= \left( \frac{m_e}{\bar{m}} n_e + \frac{m_p}{\bar{m}} n_p \right) U^\mu\,,\\
  V_p^\mu &= -\frac{m_e}{m_p} V_e^\mu.
  \label{eqn:Jep}
\end{align}
This choice also has implications for the total heat flux in the single fluid frame.
From Eq.\ \eqref{eqn:qsum} we find
\begin{align}
q^\mu \,=\,& q_e^\mu + q_p^\mu + \left(1 - \frac{m_e}{m_p}\frac{h_p}{h_e}\right) h_e V_e^\mu\,,\\
 =\,& q_e^\mu + q_p^\mu + \left(1 - \frac{\mathfrak{h}_p}{\mathfrak{h}_e}\right) h_e V_e^\mu\,,
\label{eqn:heatflux_ep}
\end{align}
where the last line is independent of the particle mass ratio, \erm{$m_e/m_p$}.
Since $h_e \sim m_e$, we find that the diffusion contribution is on the order of the electron mass,
while the total energy of the single fluid, scales like $h \sim m \approx m_p$.
Unless very large diffusion currents are present, the diffusive 
heat flux will be suppressed by the electron-ion mass ratio.\\
On the other hand, the diffusion part of the total electric current \eqref{eqn:Jsplit},
\begin{align}
    \mathcal{J}^\mu = \mathfrak{q}_e \left[\left(Z n_p - n_e\right) U^\mu - \left(1 + Z \frac{m_e}{m_p}\right) V_e^\mu\right],
    \label{eqn:current_ep}
\end{align}
is entirely dominated by the diffusion of electrons in the single fluid frame, as
$\corr{n_e \simeq Z n_p}$ in a charge-neutral collisional equilibrium.

Since the effective single fluid frame and the ion frame now almost
coincide we can, to a very good approximation,  truncate our expressions to linear order in $\left( m_e/m_p \right)$. That is
\begin{align}
  n_p \approx \,& \tilde{n}_p \label{eqn:trafo1}\,,\\
  e_p\approx \,& \tilde{e}_p \corr{-} \frac{2}{n_p}\frac{m_e}{m_p} V^\mu_e \tilde{q}_{p\,
  \mu} \,,\\
  \corr{\mathcal{Q}}_p^\mu \approx \, & \Delta^\mu_\nu \tilde{q}^\nu_p -
  \frac{m_e}{m_p}\frac{1}{n_p} V_{e\, \lambda} \Delta^\mu_\nu
  \tilde{\pi}^{\nu\lambda}_p - \left[ \tilde{h}_p + \frac{1}{n_p}
  \tilde{\Pi}_p \right] \frac{m_e}{m_p} V_e^\mu\,,\\
  \pi^{\mu\nu}_p \approx \,& \Delta^{\mu\nu}_{\alpha\beta}
  \left[\tilde{\pi}_p^{ \alpha\beta} - \corr{2}\frac{m_e}{m_p}
  \frac{1}{n_p}\tilde{q}_p^\alpha V_e^\beta  \right].
  \label{eqn:trafon}
\end{align}
Here we have used $\tilde{A}_X$ to denote quantities in the fluid rest frame of species $X$. For a detailed discussion see Appendix \ref{app:transform}.
As expected, the quantities in the ion fluid frame and in the single
fluid frame almost coincide. The small discrepancies scale exclusively with
the electron number diffusion current $V_e^\mu$.
It is, hence, crucial to understand how the evolution of the diffusion
current progresses relative to the evolution of the ion system. 

To this end, what remains to be specified are internal closure relations for the plasmas.
In what follows, we will assume that electrons and ions only interact via electromagnetic fields and, potentially, via the collision term $\mathcal{C}_e^\nu $in the joint frame, which was introduced in Eq.\ \eqref{eqn:collisions}.
In practical terms, this implies that the remaining equations can be closed for each species individually.
That is, we need to specify relations for the single fluid dissipative quantities 
$\tilde{\Pi}_X, \tilde{q}_X^\mu,\,\text{and}\ \tilde{\pi}_X^{\mu\nu}$ in the component fluid rest-frames.
These will depend on the physical state of the system. In particular, we will present two systems describing, in particular, collisionless, weakly collisional, and highly collisional electron-ion plasmas.

\subsection{Collisionless electron-ion plasmas}

In the case of a collisionless two-fluid plasma we can solve 
the local relaxation closure equations, \eqref{eqn:big_pi_less_col}, \eqref{eqn:q_less_col}, and \eqref{eqn:small_pi_less_col},  in the individual fluid frames.
For completeness we repeat these equations below.

\begin{dmath}
  \tilde{u}_X^\alpha \nabla_\alpha \left(\tilde{P}_X+\tilde{\Pi}_X\right) =
  \corr{- \frac{2}{3}} \nabla_\mu \tilde{q}_X^\mu
  -  \frac{\corr{5}}{3}\tilde{\theta}_X \left(\tilde{P}_X+\tilde{\Pi}_X\right)
  -\frac{2}{3} \tilde{q}_X^\mu \tilde{a}_{X\, \mu}
-\frac{2}{3} \tilde{\sigma}_{X\,\mu\nu} \tilde{\pi}_X^{\mu\nu}  \,,
\label{eqn:big_pi_less_col2}
\end{dmath}

 \begin{dmath}
   \tilde{u}_X^\alpha \nabla_\alpha \tilde{q}_X^{<\mu>} = \corr{+} \frac{\corr{\overline{\mathfrak{q}}_X}}{m_X} \left(\tilde{e}_X -m_X \tilde{n}_X  + \tilde{P}_{\corr{X}}+ \tilde{\Pi}_X \right) \tilde{e}_X^{\mu}
   \corr{+} \frac{\corr{\overline{\mathfrak{q}}_X}}{m_X} \tilde{e}_{X\,\alpha} \tilde{\pi}_X^{\alpha\mu}
   \corr{+} \frac{\corr{\overline{\mathfrak{q}}_X}}{m_X} \tilde{b}_X^{\mu\alpha} \tilde{q}_{X\,\alpha}
-
\left(\tilde{e}_X-m_X \tilde{n}_X + \tilde{P}_X+\tilde{\Pi}_X\right) \tilde{u}_X^\alpha\nabla_\alpha \tilde{u}_X^{\mu} 
- \tilde{\theta}_X \tilde{q}_X^{\mu} 
 \corr{-} \tilde{a}_{X\,\alpha} \tilde{\pi}_X^{\mu\alpha} - \tilde{q}_X^{\alpha}\nabla_\alpha \tilde{u}_X^\mu -\frac{1}{\tau_q^{\corr{X}}} \tilde{q}_X^\mu\,,
\label{eqn:q_less_col2}
\end{dmath}

\begin{dmath}
  \tilde{u}_X^\alpha \nabla_\alpha \tilde{\pi}_X^{<\mu\nu>} =
   + 2 \frac{\corr{\overline{\mathfrak{q}}_X}}{m_X}
   \tilde{b}^{<\mu}_{X\ \ \alpha} \tilde{\pi}^{\nu>\corr{\alpha}}_{X} - \tilde{\theta}_X \tilde{\pi}_X^{\mu\nu}
-2 \tilde{a}_X^{<\mu} \tilde{q}_X^{\nu>}
-2 \tilde{\pi}_X^{\alpha <\mu}\nabla_\alpha \tilde{u}_X^{\nu>} 
\corr{- 2}\left(\tilde{P}_X+\tilde{\Pi}_X\right) \tilde{\sigma}_X^{\corr{\mu\nu}}
-\frac{1}{\tau_\pi^{\corr{X}}} \tilde{\pi}_{\corr{X}}^{\mu\nu}\,.
\label{eqn:small_pi_less_col2}
\end{dmath}
{
Solving these equations provides \corr{a local covariant 14-moment approximation
to collisionless plasmas in curved spacetimes within the species-wise
low-temperature regime stated above}.
In particular, this description retains all electron inertia terms, resistive and Hall terms. Moreover, all anisotropic pressure contributions are retained, with all transport coefficients apart from the relaxation times being fixed, in principle, by first-principle kinetic theory calculations.
}
\subsection{Weakly collisional systems}
\label{sec:weakly_col}

{For some black hole accretion problems collisions become important \citep{Ressler:2015ipa}.} If we want to account for Coulomb collisions inside the plasma (i.e. electron-electron) collisions, we need to include effective collisionality in these equations.
Such a situation might arise inside accretion disks around supermassive black holes, 
where electron heat fluxes and anisotropic viscosities would affect the flow structure and 
electron heating.

Different from the previous case, we proceed by introducing an equilibrium state for each fluid.
That is, we specify equations of state for the pressures $\tilde{P}_X$. 
A sensible, and simple choice, would be a simple gamma-law as is commonly done in black hole accretion \citep{EventHorizonTelescope:2019pcy},
\begin{align}
    \corr{\tilde{P}_e} &= \tilde{n}_e \tilde{T}_e = m_e \tilde{n}_e \tilde{\varepsilon}_e \left(\gamma_e-1\right)\,,\\
    \corr{\tilde{P}_p} &= \tilde{n}_p \tilde{T}_p = m_p \tilde{n}_p \tilde{\varepsilon}_p \left(\gamma_p-1\right)\,,
\end{align}
where $\gamma_e$ and $\gamma_p$ are the adiabatic coefficients of the plasmas.\\

For these equations of state, we can solve Eq.\ \eqref{eqn:eX} for the
temperature $\tilde{T}_X$ of each species,
\begin{align}
  \tilde{T}_X = \left[ E_X\sqrt{n_X^2 -V_X^2}  - m_X n_X^2 \right] \frac{
    \tilde{\Gamma}_X-1}{n_X^2 + \left( \tilde{\Gamma}_X-1 \right) V_X^2},
  \label{eqn:Tx}
\end{align}
where
\begin{align}
  E_X =  e_X  \corr{-} \frac{2}{\sqrt{n_X^2 -V_X^2}} V_X^\mu \tilde{q}_X^\nu
  g_{\mu\nu} - \frac{1}{\corr{n_X^2}} V_X^\mu V_X^\nu
  \tilde{\pi}_{X\,\mu\nu}.
  \label{eqn:TxE}
\end{align}
Within the mass hierarchy adopted in this section, the joint-frame total
isotropic pressure per ion reduces to
\begin{align}
  \corr{T_{p,{\rm iso}} \equiv \frac{P_p+\Pi_p}{n_p}}
  =  \tilde{T}_p - \frac{m_e}{m_p}\corr{\frac{2}{3}}
  \frac{V_e^\mu \corr{\tilde{q}}_{p\,\mu}}{n_p^2}.
  \label{eqn:Tp_approx}
\end{align}
We can see that in the absence of intrinsic ion heat fluxes, the ion
temperature in the ion frame and in the single fluid frame begin to
coincide. 
We can similarly recover the electron temperature from Eq.\ \eqref{eqn:Tx},
however, no simplification a priori is possible.

In the following we will consider the case of ion-ion and electron-electron
collisions. This scenario has first been considered by \citet{1965RvPP....1..205B}
in the non-relativistic case. 
Additionally, relative diffusion between electrons and ions is possible due
to the coupling the electromagnetic field in Eq.\ \eqref{eqn:collisions}.
Following \citet{kulsrud2020plasma}, we assume that ion-ion collisions lead to an effective
anisotropic shear stress. In the limit of vanishing Larmor radius, this
will have to approach the Braginskii limit \citep{1965RvPP....1..205B}.
Following \cite{Most:2021oha}, this can be achieved by imposing the
following closure relations of the ion pressure tensor, see also \citet{Denicol:2018rbw},
\begin{dmath}
    \tilde{u}_p^\alpha \nabla_\alpha \tilde{\pi}_p^{<\mu\nu>} =
    -2\frac{\bar{\nu}_p}{\tau^p_\pi} \nabla^{< \mu } \tilde{u}_p^{\nu >}  +  \delta^p_{\pi B}
    \tilde{b}^{<\mu}_{p\,\alpha} \tilde{\pi}_p^{\nu>\alpha} 
    -\frac{1}{\corr{\tau}_{\pi}^p} \tilde{\pi}_p^{\mu\nu} \,,
\end{dmath}
\begin{dmath}
\tilde{\pi}_e^{\mu\nu}=0\,.
\end{dmath}
Additionally, we assume that intra-species collisions will drive a heat
flux. Again, inspired by \cite{Most:2021oha}, we can assume that to within
first-order this leads to a closure of the following form,
\begin{dmath}
   \tilde{u}_p^\alpha \nabla_\alpha \tilde{q}_p^{<\mu>} =-
   \frac{\corr{1}}{\tau_{q}^p}\tilde{q}_p^\mu  - \erm{\corr{\frac{\kappa_p}{\tau_q^p}}}
   \nabla^{<\mu>} \tilde{T}_p - \erm{\corr{\frac{\kappa_p}{\tau_q^p}}} \tilde{T}_p
   \tilde{u}_p^\alpha\nabla_\alpha \tilde{u}_p^\mu + \delta^p_{q} \tilde{b}_p^{\mu\nu} \tilde{q}_{p\,\nu}   \,.
\end{dmath}
\begin{dmath}
   \tilde{u}_e^\alpha \nabla_\alpha \tilde{q}_e^{<\mu>} =-
   \frac{\corr{1}}{\tau_{q}^e}\tilde{q}_e^\mu  - \erm{\corr{\frac{\kappa_e}{\tau_q^e}}}
   \nabla^{<\mu>} \tilde{T}_e - \erm{\corr{\frac{\kappa_e}{\tau_q^e}}} \tilde{T}_e
   \tilde{u}_e^\alpha\nabla_\alpha \tilde{u}_e^\mu + \delta^e_{q} \tilde{b}_e^{\mu\nu} \tilde{q}_{e\,\nu}   \,.
\end{dmath}
Here the magnetic-response coefficients are signed:
$\corr{\delta_q^p=+Z\mathfrak{q}_e/m_p}$ and
$\corr{\delta_q^e=-\mathfrak{q}_e/m_e}$.
We further assume that bulk scalar pressures in their respective local frames
vanish
\begin{align}
  \corr{\tilde{\Pi}}_X =0\,,
  \label{eqn:Pivanish}
\end{align}
{since we do not take into account bulk viscosities.}
Coulomb collisions arising from electron-electron and ion-ion interactions give
rise to effective transport coefficients. These have been derived in the weakly collisional limit of large gyrofrequencies, $\Omega_e= \mathfrak{q}_e B/m_e $ and $\Omega_p= \corr{\mathfrak{q}_p} B/m_p \corr{= Z\mathfrak{q}_e B/m_p}$. A list of all transport coefficients can be found, e.g. in \citet{kulsrud2020plasma}.

\section{Dissipative magnetohydrodynamics for resistive relativistic single-fluid plasmas}
\label{sec:dissMHD}
So far we have considered closures that explicitly keep all two-fluid
degrees of freedom.  
Current state-of-the-art simulations of ideal and
resistive relativistic magnetohydrodynamics, however, normally treat
only single (ion-)fluid plasmas (see \citealt{marti2015grid} for a review).  In the following, we would like to derive this
limit from the two-fluid equations considered in this work.  

We start by considering the single fluid plasma described by Eq.\
\eqref{eqn:Tmunutot}.  We now neglect also all terms linear in
$m_e/m_p$ in
Eqs.\ \eqref{eqn:trafo1}-\eqref{eqn:trafon}, that is
\begin{align}
  n_p \approx \tilde{n}_p \,,~
  n \approx n_p \,,~
  e_p\approx \tilde{e}_p \,,~
  e \approx  e_p\,, ~ \nonumber\\
  q_p^\mu \approx \Delta^\mu_\nu \tilde{q}^\nu_p\,,~
  \pi^{\mu\nu}_p \approx \Delta^{\mu\nu}_{\alpha\beta}
  \tilde{\pi}_p^{ \alpha\beta}.
\end{align}
Consistent with the assumption of a single ion-fluid we also neglect anisotropic
electron pressures,
\begin{align}
    \pi_e^{\mu\nu} \approx 0,
\end{align}
but keep the electron heat flux, in line with the discussion of
weakly collisional plasmas in Sec.\ \ref{sec:weakly_col}.

If we neglect electron shear viscosity and heat conduction, to lowest order
in $m_e/m_p$, the electric current reduces to
\begin{align}
    \mathcal{J}^\mu \approx \mathfrak{q}_e \left[\left(Z n - n_e\right) U^\mu - V_e^\mu\right],
    \label{eqn:current_ep2}
\end{align}
which will be given in terms of the electron Ohm's law. The heat flux \eqref{eqn:heatflux_ep} remains unchanged.  Furthermore, if we
neglect all direct couplings between the electron \erm{energy diffusion vector}
$\mathcal{Q}_e^\mu$ and the single fluid velocity $U^\nu$ and its
gradients, we arrive at
  \begin{dmath}
    U^\lambda \nabla_\lambda \mathcal{Q}_e^{<\mu>} =
    -\mathfrak{q}_e n_e e^\mu \corr{-} \mathfrak{q}_e b^{\mu\nu} \erm{\corr{V_{e\, \nu}}} -\nabla^{<\mu>} P_e - \corr{\left(e_e+P_e\right)} U^\lambda \nabla_\lambda U^\mu - \frac{1}{\tau} \mathcal{Q}_e^{\mu}.
    \label{eqn:emomentum_ep}
  \end{dmath}
At this point, it remains to specify the electron enthalpy, which
effectively amounts to fixing the electron temperature.  This could either
be done by evolving the electron energy or by algebraically fixing the
electron temperature $T_e$. The latter is commonly done in the post-processing of black-hole accretion simulations, where the electron temperature governs part of the emission process (e.g., \citealt{Ressler:2015ipa}).
Different from such ad-hoc assumptions, the explicit inclusion of dissipative terms allows us to explicitly track the change in electron temperature in a self-consistent manner.
More specifically, using the same assumption
as for the electron momentum equation \eqref{eqn:emomentum_ep} we find,
\begin{align}
    \nabla_\mu \left(e_e U^\mu \right) = - P_e \nabla_\mu U^\mu \corr{-}  \nabla_\mu \erm{\mathcal{Q}_e^\mu} - \mathfrak{q}_e e_\mu V_e^\mu.
    \label{eqn:e_evol_ep}
\end{align}
We can now see that in this minimalistic model the electron energy only changes via diffusion, Ohmic
heating, and adiabatic compression.
Furthermore, since we retain ion heat fluxes and anisotropic stresses,
these can be closed using the 14-moment relations given in Sec.
\ref{sec:col_closure}. Keeping only terms that are first-order in gradients,
we arrive at
\begin{align}
   U^\mu \nabla_\mu q_p^{\left<\corr{\mu}\right>} =& - \corr{\frac{\kappa_p}{\tau_q^p}} \nabla^{\left<\mu\right>} \tilde{T}_p - \corr{\frac{\kappa_p}{\tau_q^p}} \tilde{T}_p U^\alpha\nabla_\alpha U^\mu \nonumber \\
   &\corr{+ \delta^p_{q}} b^{\mu\nu} q_{p\, \nu} - \frac{1}{\tau_q^p} q_p^{\corr{\mu}}\,, \\
   \tilde{u}_e^\alpha \nabla_\alpha \tilde{q}_e^{\left<\mu\right>} =&- \corr{\frac{\kappa_e}{\tau_q^e}} \nabla^{\left<\mu\right>} \tilde{T}_e - \corr{\frac{\kappa_e}{\tau_q^e}} \tilde{T}_e \tilde{u}_e^\alpha\nabla_\alpha \tilde{u}_e^\mu \nonumber \\
   &\corr{+} \delta^e_{q} b^{\mu\nu} \tilde{q}_{e\, \nu} - \frac{\corr{1}}{\tau_{q}^e}\tilde{q}_e^\mu  \,, \label{eqn:eheatflux_f}\\
   U^\alpha \nabla_\alpha \pi_p^{\left<\mu\nu\right>} =&
    -2\frac{\bar{\nu}_p}{\corr{\tau^p_\pi}} \nabla^{\left< \mu\right.} U^{\left. \nu \right>} -\frac{1}{\corr{\tau^p_\pi}} \pi^{\mu\nu}_p\nonumber\\
    &+ \delta^p_{\pi B} b^{\alpha\beta}\Delta^{\mu\nu}_{\alpha \kappa} g_{\lambda \beta}\pi_p^{\kappa\lambda}
\end{align}

Together with Eqs. \eqref{eqn:emomentum_ep} and \eqref{eqn:e_evol_ep}, the
system described here is \corr{a dissipative MHD model} for a resistive single
component plasma with dissipative corrections from the secondary fluid.
Crucially, for consistency with charge neutrality, a minimal number of degrees of freedom of the second species needs to be retained. In particular these are the electron number density and momentum. In addition, we also retain electron temperature evolution, which provides a consistent way to extract electron temperatures beyond previously adopted ad-hoc approaches \citep{Ressler:2015ipa,Chael:2018aeq}.

\subsection{Single-fluid dissipative magnetohydrodynamics}

Instead of solving the electron energy equation \eqref{eqn:e_evol_ep}, one could alternatively specify the electron temperature via an effective prescription for the electron specific enthalpy
\begin{align}
    \mathfrak{h}_e = \mathfrak{h}_e \left(\mathfrak{h}_p, n, e \right).
\end{align}
Such relations have been proposed in \citet{Howes:2010pr,Kawazura:2018vsm,Rowan:2017cao}.
A simple choice, commonly done for single fluid plasmas is to neglect the electron temperature all together. In that limit, we may simply set $\mathfrak{h}_e \approx 1$, assuming that thermal contributions are negligible compared to the rest-mass energy.
It is important to understand that under this assumption the electron temperature can no longer be recovered from the evolution system. On the other hand, such a simplification allows us to rewrite Eq.\ \eqref{eqn:emomentum_ep} to read,
\begin{align}
    U^\lambda \nabla_\lambda V_e^{\left<\mu\right>} =
    -\frac{\mathfrak{q}_e}{m_e} n_e e^\mu \corr{-} \frac{\mathfrak{q}_e}{m_e} b^{\mu\nu} V_{e\, \nu} - \frac{1}{\tau} V_e^{\mu}.
    \label{eqn:emomentum_ep_reduced}
\end{align}
{This is the most minimalistic form of a self-consistent electron momentum equation. In this case, both electron inertia terms, as well as resistivity and the Hall term, are kept.}
Together with the conservation of electric charge \eqref{eqn:current_ep}
and \eqref{eqn:Jcons}, this completes the electron closure relations. 
{If we further neglect electron inertia terms on the LHS,
we arrive at the standard Ohm's law
}
\begin{align}
   V_e^\mu \approx \corr{-}\sigma e^\mu - \tau \frac{\mathfrak{q}_e}{m_e} b^{\mu\nu} V_{e\, \nu},
    \label{eqn:emomentum_ep_reduced_no_evol}
\end{align}
where $\sigma = \tau n_e \frac{\mathfrak{q}_e}{m_e}$ is the particle-diffusion
response coefficient and
$\corr{\sigma_E=\mathfrak{q}_e\sigma=\tau n_e\mathfrak{q}_e^2/m_e}$ is the
electric conductivity.
The second term will give rise to a Hall effect with varying
degree of anisotropy.  It is important to note that
dropping the advection operator in Eq.\ \eqref{eqn:emomentum_ep_reduced}
breaks the strong hyperbolicity of the system, as has been shown for previously used resistive relativistic MHD systems \citep{Schoepe:2017cvt}. 
Hence, the consistent evolution of $V_e^\mu$, see Eq. \eqref{eqn:emomentum_ep_reduced}, is crucial to the causality and stability of the dissipative MHD system.

\subsection{Force-free limit}

When modeling compact object magnetospheres the copious production of electron-positron pairs
will lead to an efficient screening of the electric field component parallel to the magnetic field \citep{Goldreich:1969sb}. This limit has been extensively studied in global magnetospheric models ( e.g., \citealt{Spitkovsky:2006np,Alic:2012df,Palenzuela:2012my,Parfrey:2013gza,Carrasco:2018kdv,Most:2020ami}),
although one is not always able to meaningfully capture reconnection physics in current sheets (e.g., \citealt{Mahlmann:2020yxn,Ripperda:2021pzt}). 
In the following, we will give a brief outline of the force-free limit in the dissipative MHD system.
As we will show, this system naturally contains the force-free limit, where the transition between the two is not ad-hoc \citep{Palenzuela:2012my} but physically motivated in terms of conductivities and gyration frequencies.

In this force-free electrodynamics limit, the Lorentz force will vanish, i.e.
\begin{align}
  \nabla_\mu T_{\rm hydro}^{\mu\nu} = \corr{+} F^{\nu\mu} \mathcal{J}_\mu \approx 0,
  \label{eqn:TFFFE}
\end{align}
where the electromagnetic fields act upon the fluid via
Lorentz forces, viz.
\begin{align}
  F^{\mu\nu} \mathcal{J}_\nu = \left( q_u g^{\mu\nu} \corr{- \mathfrak{q}_e} U^\mu V_e^\nu
  \right)e_\nu \corr{- \mathfrak{q}_e} b^{\mu\nu}V_{e\,, \nu}.
  \label{eqn:FJFFE}
\end{align}
In writing the above we have made use of the short-hand $q_u = \mathfrak{q}_e \left(Z n - n_e\right)$.
In order to obtain the force-free limit the Lorentz force
\begin{align}
    F^{\mu\nu} \mathcal{J}_\nu \approx 0\,,
\end{align}
 needs to vanish. From Eq.\ \eqref{eqn:FJFFE} this can generally only be achieved if 
 the comoving electric field and the diffusion current perpendicular to the (comoving) magnetic field vanish simultaneously. That is
\begin{align}
    & e_\mu\, \approx\, 0\,,\\
    & b_{\mu\nu} V_e^\nu\, \approx\, 0\,.
\end{align}
The latter condition directly implies that
\begin{align}
    V_e^\mu \approx V_\parallel \frac{1}{\sqrt{b^2}}b^\mu\,,
\end{align}
where $V_\parallel$ is the diffusion current parallel to the comoving  magnetic field $b^\mu$. As such, the diffusion current will naturally maintain
the ${^\ast}F^{\mu\nu} F_{\mu\nu}=0$ condition, see also the discussions in \citet{Palenzuela:2012my} and \citet{Paschalidis:2013gma}.
From Eq.\ \eqref{eqn:emomentum_ep_reduced}, we can now see that the equations presented in this work naturally recover the force-free limit when both the electrical conductivity $\sigma$ and gyration frequency $\Omega_e$ are large, as the diffusion current $V_e^\mu$ needs to remain finite. Since $\Omega_e \sim \sqrt{b^2}$, in this limit magnetic field strength is large. 
Hence, this system of equations naturally contains the force-free limit and can also achieve it self-consistently in the limit of a strongly magnetized perfectly conducting plasma.

In other words, the system presented here is well suited to study neutron star magnetospheres with self-consistent non-ideal effects, allowing to accurately capture reconnection processes,\corr{though it will be left to future work to establish actual well-posedness for this system of equations.}

\subsection{Non-resistive dissipative magnetohydrodynamics}

As a final reduction, we can further assume that conductivity is infinite $\sigma\rightarrow \infty$. In this case, the comoving electric field $e^\mu$ vanishes.
Since the out-of-frame diffusion current  has no effective source-term in Eq.\ \eqref{eqn:emomentum_ep_reduced}, any initial condition in $V_e$ will only be (anisotropically) advected and decay on a time scale $\tau$. Hence, we may then set
\begin{align}
    V_e^\mu =0.
\end{align}
Because of the divergence constraint on the electric field, 
\begin{align}
    U_\mu \nabla_\nu F^{\mu\nu} = 4\pi U_\mu \mathcal{J}^\mu\,,
\end{align}
this choice uniquely fixes the electron number density via \eqref{eqn:maxwell},
\begin{align}
  n_e = Z n  - \frac{1}{4\pi \mathfrak{q}_e}\left[ \nabla_\mu e^\mu \corr{- e^\mu U^\lambda\nabla_\lambda U_\mu} -  b^{\mu\nu} \nabla_\mu U_\nu\right].
\end{align}
Since the electron number density and temperature are now fixed, we can drop the evolution equation for the electron \corr{diffusion current} \eqref{eqn:emomentum_ep_reduced}.

At this point, all electron degrees of freedom have been stripped from the system, and we are left with a single perfectly conducting ion fluid.
We then only retain first-order out-of-equilibrium corrections of the ion fluid,
\begin{align}
   U^{\corr{\alpha}} \nabla_{\corr{\alpha}} q^{\left<\mu\right>} =& - \frac{\kappa}{\tau_q} \nabla^{\left<\mu\right>} T - \frac{\kappa}{\tau_q} T U^\alpha\nabla_\alpha U^\mu + \delta_{q} b^{\mu\nu} q_{\nu} - \frac{1}{\tau_q} q^\mu\,, \\
   U^\alpha \nabla_\alpha \pi^{<\mu\nu>} =&
    -2\frac{\bar{\nu}}{\tau_\pi} \nabla^{\left< \mu\right.} U^{\left. \nu \right>} -\frac{1}{\tau_\pi} \pi^{\mu\nu}+ \delta_{\pi B} b^{\alpha\beta}\Delta^{\mu\nu}_{\alpha \kappa} g_{\lambda \beta}\pi^{\kappa\lambda}\,,
\end{align}
where for simplicity we have dropped all subscripts, as only one fluid is present.
These equations now describe the evolution of a dissipative single fluid coupled to magnetic fields in the non-resistive limit \citep{Denicol:2018rbw}, which have recently been studied by \citet{Most:2021oha}.

\section{Conclusions}

In this work we have systematically investigated the description of a
general-relativistic plasma consisting of two charged components
interacting through electromagnetic fields. To include non-ideal out-of-equilibrium effects, we have gone beyond the traditional
magnetohydrodynamical approach, and have adopted a 14-moment formulation
\citep{Denicol:2012cn,Denicol:2018rbw,Denicol:2019iyh}, 
which is a generalization of Grad's 13-moment approach in
Newtonian plasma physics. As a moment expansion of the general-relativistic 
Boltzmann equation, such an approach always requires the specification of a
closure relation, expressing higher moments in terms of the main 14-moments
used here. Building on closure relations devised in the context of nuclear
physics applications, we present the general 14-moment decomposition of the
Boltzmann equation in the absence of collisions.
By comparing these equations to the general form of a collisional closure
in the presence of electromagnetic fields \citep{Denicol:2019iyh}, we show that these
equations are consistent with the collisional limit, except for the
presence of additional heat flux terms. Extending a similar closure from  Newtonian
10-moment descriptions of collisionless plasmas \citep{2015PhPl...22a2108W}, we propose a simple closure
relation for the collisionless plasma. Although formally collisional in
nature, such a closure has compared favourably to full numerical solutions
of the Newtonian Vlasov equation \citep{2015PhPl...22a2108W}.
{\corr{The expansion used to derive this closure is controlled only when each
species satisfies $T_X/m_X\ll1$, with relative closure errors of order
$T_X/m_X$. Consequently, this particular closed system is not applicable to
relativistically hot electrons or pair plasmas; those regimes require a
different closure or resummation of the exact moment hierarchy (see
\citealt{Tinti:2018qfb} for a discussion).}}

In order to recast the system into a form typically adopted in the
numerical study of relativistic plasmas \citep{Palenzuela:2008sf,Ripperda:2019lsi},
we rewrite the 
two-fluid equations as a single fluid system with potentially large
out-of-equilibrium corrections. While being exactly equivalent to modeling
two separate general fluids, this description is particularly useful for the
case of two-component plasmas with large particle mass ratios. In this case 
the system naturally takes the form of a single fluid with dissipative
corrections. These split into internal out-of-equilibrium processes
(e.g. electron-electron, ion-ion collisions) as well as inter-species
contributions (electron-ion). Remarkably, the latter part takes the form of
a 14-moment (collisional) closure. 
This fundamentally illustrates that the entire out-of-equilibrium sector
can be solved using recently developed methods for 14-moment dissipative
relativistic MHD \citep{Most:2021oha}.

Another feature of recasting the two-fluid system in dissipative single-fluid
form is the ability to systematically approximate and simplify the dissipative
sector. By expanding all expressions in the particle mass ratio, we are able
to derive an effective Ohm's law including only first-order dissipation and
anisotropy effects. This most general form of dissipative MHD retains a
minimal amount of electron degrees of freedom. In particular, it allows for
the consistent evolution of electron temperature, which is driven by
Ohmic and viscous heating. Such an approach is potentially relevant for
current simulations of black-hole accretion which can either model electron
temperature only in post-processing \citep{Howes:2010pr,Rowan:2017cao,Kawazura:2018vsm}, or using approximate
calculation that rely on non-convergent grid dissipation \citep{Ressler:2015ipa}.
It is important to stress at this point that the consistent treatment of
electron momentum naturally requires a dissipative heat-flux correction to
the effective single ion-fluid. In order to allow for causal evolutions in a second-order formulation,
the heat flux cannot be algebraically related to the fluid and its
gradients \citep{Israel:1979wp}, and instead
requires separate evolution equations. Fundamentally, this implies that in
a relativistic setting the electric current entering the Maxwell equation
will require a separate evolution equation, where the time derivatives
cannot be dropped unlike in Newtonian contexts. While we postpone 
the issue of deriving strict causality and, in turn, strong hyperbolicity
 conditions for the system presented here, our results might explain why previous
approaches to resistive relativistic MHD \citep{Palenzuela:2008sf,Dionysopoulou:2012zv,Ripperda:2019lsi} have been found to be only weakly hyperbolic \citep{Schoepe:2017cvt}. Further studies are clearly needed in this subject, especially given that only very recently the causal properties of Israel-Stewart-like theories (without electromagnetic field effects) have been understood in the nonlinear regime \cite{Bemfica:2019cop,Bemfica:2020xym}.

Apart from a numerical assessment of this system and a potential comparison
with collisionless relativistic particle-in-cell simulations
\citep{2015PhPl...22a2108W,doi:10.1063/5.0012067}, several extensions seem possible.
Motivated by the need to model relativistic reconnection in neutron star
and black hole magnetospheres, we have mainly focused on a two-component
plasma that could either be electron-ion or electron-positron.
In the context of black hole magnetospheres a potential coupling to photon
radiation transport would allow to model dynamical pair processes in the jet
regions of supermassive black holes \citep{Moscibrodzka:2009gw}. 
In the context of neutron star interiors, the inclusion of weak-interaction
out-of-equilibrium effects coupled to neutrino radiation would allow for
the self-consistent study of dissipative effects in neutron star mergers
\citep{Most:2021zvc}.
Finally, neutron star interiors are believed to consist of superfluid
phases (for a review, see \citealt{Chamel:2017wwp}). Recent models seem to indicate that these can be modeled
using similar Israel-Stewart like approaches \citep{Gavassino:2021crz}.
It thus seems natural to extend this 14-moment closure to a general
multi-fluid approach (see also \citealt{Andersson:2016fva,Andersson:2016wnc}).

\section*{Acknowledgements}
The authors thank Lev Arzamasskiy, Amitava Bhattacharjee, Gabriel Denicol, Chuanfei Dong, Ammar Hakim,
James Juno, Alex Pandya, Frans Pretorius, James Stone, Jason Ten Barge and the members of the CCA-PPPL collaboration for insightful discussions and comments related to this work. 
ERM gratefully acknowledges support from postdoctoral fellowships 
at the Princeton Center for Theoretical Science, the Princeton
Gravity Initiative, and the Institute for Advanced Study. 
Research at the Flatiron Institute is supported by the Simons Foundation. 
JN is partially supported by the U.S. Department of Energy, Office of Science, Office for Nuclear Physics under Award No.\ DE-SC0021301.
\\
\section*{Data Availability}
No new data was produced in this study.
%

%
\bibliographystyle{mnras}
\bibliography{mnras_template,non_inspire} 
%
%
%
%
%

\newpage
\appendix

\section{Two-fluid transformations}
\label{app:transform}
Having derived evolution equations for the effective
single fluid frame, we need to relate the heat fluxes and anisotropic stresses described in 
Eq. \eqref{eqn:Tindiv}.
To this end, we consider the rest-frames of each fluid, for which we can
define individual closure relations, see Sec. \ref{sec:gkyell}. 

{Adopting an Eckart frame in the rest-frame of each fluid, we may write}
{
\begin{align}
  N_e^\mu = \tilde{n}_e u^\mu_e\,,
  \label{eqn:N2e_eckart}\\
  N_p^\mu = \tilde{n}_p u^\mu_p\,,
  \label{eqn:N2p_eckart}
\end{align}
}
{where we have introduced particle number densities $\tilde{n}_{e/p}$ and velocities
$u_{e/p}^\mu$.}
{We can then relate these number densities to the one seen in the single fluid frame via}
\begin{align}
  \tilde{n}_X^2 = n_X^2 - g_{\mu\nu} V^\mu_X V^\nu_X,
  \label{eqn:ntile_from_n}
\end{align}
where $X$ denotes either electrons $\left( e \right)$ or ions/{positrons} $\left( p
\right)$.
This implies that the number density as seen by an observer comoving with
the single-fluid frame, will be different from the one in the component
fluids rest-frame. This difference is given by the diffusion current.
In the following, we will denote all quantities in the electron or ion frame with as $\tilde{A}_X$.
Similarly, we can relate the energy density and pressure in the two frames.
From Eq.\ \eqref{eqn:Tfsplit} we then find that
\begin{dmath}
  e_X = \tilde{e}_X \left[ 1 + \frac{V_X^2}{n_X^2 - V_X^2} \right] +
  \left( \tilde{P}_X + \tilde{\Pi}_X \right) \frac{V_X^2}{n_X^2 - V_X^2}
  \corr{+} \frac{2}{\tilde{n}_X} V_X^\mu \tilde{q}_X^{\nu} g_{\mu\nu} +
  \frac{1}{\corr{n_X^2}} V^\mu_X V^\nu_X \tilde{\pi}_{X\, \mu\nu}\,,
  \label{eqn:eX}
\end{dmath}
\begin{dmath}
  P_X + \Pi_X  = \tilde{P}_X  + \tilde{\Pi}_X + \frac{1}{3} \left( e_X - \tilde{e}_X \right). \label{eqn:pX}
\end{dmath}

It can be seen that also the enthalpy receives (multiplicative) corrections
related to the magnitude of the diffusion current $V_X^\mu$.
 Similarly, we can write
\begin{dmath}
  \corr{\mathcal{Q}}_X^\mu = \frac{n_X}{\tilde{n}_X} \Delta_\nu^\mu \tilde{q}^\nu_X  +
  \corr{\frac{1}{n_X}}\Delta^\mu_\nu \tilde{\pi}^{\nu\corr{\lambda}}_X \corr{V_{X\,\lambda}}
   +\left[ \frac{1}{n_X} \tilde{q}_X^\nu V^\lambda_X g_{\lambda\nu} \corr{+}
  \frac{\tilde{e}_X + \tilde{P}_X + \tilde{\Pi}_X}{\tilde{n}_X} n_X
\right]\frac{1}{\tilde{n}_X} V^\mu_X. 
  \label{eqn:qfromtilde}
\end{dmath}
\begin{dmath}
\pi^{\mu\nu}_X = \Delta^{\mu}_\alpha \Delta^\nu_\beta \tilde{\pi}^{\alpha\beta}  +
\frac{\tilde{e}_X + \tilde{P}_X + \tilde{\Pi}_X}{\tilde{n}^2_X} V^\mu_X
V^\nu_X  + \frac{1}{\tilde{n}_X} \tilde{q}^\alpha_X \left[ V^\mu_X
\Delta^\nu_\alpha + \Delta^\mu_\alpha V^\nu_X\right]
- \frac{1}{3}\Delta^{\mu\nu} \left[ \frac{\left( \tilde{e}_X + \tilde{P}_X
\corr{+ \tilde{\Pi}_X}\right)}{\tilde{n}_X^2} g_{\alpha\beta}V_X^\alpha V_X^\beta + 2
\frac{1}{\tilde{n}_X} g_{\alpha\beta} \tilde{q}_{X}^{\alpha} V^\beta_X +
\frac{1}{\corr{n_X^2}} V^\alpha_X V^\beta_X \tilde{\pi}_{X\, \alpha\beta} \right]
. \label{eqn:pifromtilde}
\end{dmath}


Using those relations, we can easily translate between the individual
component frames and the joint single fluid frame.
This turns out to be particularly useful when relating individual
component fluid
dissipation (e.g. ion-ion collisions) to the joint single fluid frame.

\label{lastpage}
\end{document}